\documentclass[twocolumn,english,aps,superscriptaddress,pre,longbibliography]{revtex4-1}

\usepackage[T1]{fontenc}
\usepackage[latin9]{inputenc}
\usepackage{color}
\usepackage{babel}
\usepackage{amsmath}
\usepackage{amssymb}
\usepackage{graphicx}
\usepackage{esint}
\usepackage[unicode=true,pdfusetitle,
 bookmarks=true,bookmarksnumbered=false,bookmarksopen=false,
 breaklinks=true,pdfborder={0 0 0},backref=false,colorlinks=true]
 {hyperref}
\hypersetup{
 linkcolor=blue,citecolor=blue,urlcolor=blue}

\makeatletter

\usepackage{pslatex}

\usepackage{color}
\usepackage{babel}
\usepackage{subfigure}
\makeatother

\begin{document}

\title{Scaling in Local Optimal Paths Cracks}

\author{Aurelio W. T. de Noronha}
\author{Levi R. Leite}
\affiliation{Instituto de Ciências Exatas e Naturais, Universidade da Integração Internacional da Lusofonia Afro-Brasileira, 62790-970 Redenção, Ceará, Brazil }


\date{\today}
\begin{abstract}
How local cracks can contribute to the global cracks landscape is a 
goal of several scientific topics, for example, how bottlenecks can 
impact the robustness of traffic into a city?  
In one direction, cracks from cascading failures into networks were generated using 
a modified Optimal Path-Cracking (OPC) model proposed by Andrade et al \cite{Andrade2009}.
In this model, we broke links of maximum energies from optimal paths between two 
sites with internal (euclidean) distances $l$ in networks with linear size $L$.
Each link of this network has an energy value that scales with a power-law 
that can be controlled using a parameter of the disorder $\beta$.   
Using finite-size scaling and the exponents from percolation 
theory we found that the mass of the cracked links on local optimal paths 
scales with a power-law $l^{0.4}$ as a separable equation from $L$ and
that can be independent of the disorder parameter.

\end{abstract}
\keywords{local scaling effects}

\maketitle

\section{Introduction}

Microcracks can be found in several materials and may diffuse until fracture this material. 
Modeling the material as a network with several nodes and links and microcracks as
nodes or links cracked locally, we can be studying the evolution of the 
microcracks until fracture the entire network using
scaling theory. 
Here, materials can be a network of a traffic of
a city, circuits, an electrical grid, etc. And microcrack can be
a local crack as a bottleneck of traffic, a burnt resistor, a burnt
transmission line, etc. Some phenomena could be mapped as
networks with nodes, links, and properties, that can be studied
using scaling theory ~\cite{Family1985,Barabasi1999,Barabasi2001,Dorogovtsev2002,Newman2002,Xu2007,Estrada2012}.  

The link of a network can have a cost, for example, a time travel cost from a start node to an ending node. 
The set of all links with a time travel cost can be treated as network of time travel of a city. 
Choosing two points (nodes) of a city (network) to build a path, probably this path is the path with
the lowest time travel cost (optimal path). 
Finding the optimal path between two points in a disordered system is a very important goal of study for science and technology since a long
time ago ~\cite{Kirkpatrick1985,Kardar1986,Kardar1987,Perlsman1992,Cieplak1994,Cieplak1995,Cieplak1996,Schwartz1998,Porto1999,Buldyrev2004,Buldyrev2006}.
This optimization problem is present in our daily lives
when, for example, we make use of the global positioning
system (GPS) to trace the best route to arrive at our destination or 
we choose the optimal path from house to job.
The characterization of the optimal path is highly relevant to study fractures
~\cite{Andrade2009,Oliveira2011}, 
polymers in random environments~\cite{Schwartz1998},
cascading failures~\cite{ZHENG2007700,Martin2016}, 
transport in porous media~\cite{Stanley1999} 
and robustness of complex networks~\cite{Albert2000,Holme2002,Moreno2003,Ellinas2015}.
When the optimal paths are broken in the network, the robustness is compromised. 
The robustness of networks has been studied with different algorithm 
strategies, for example, the geometrical properties of complex networks~\cite{PaoloMasucci2016}, attacks with the removal of sites~\cite{Parshani2010b,Buldyrev2010a,Gao2011} or the 
removal of links~\cite{Carmona2020}.

This paper aims to study how the process of finding a new optimal path 
when the current path is broken evolves through time until reaches 
its final configuration.
When optimal paths are broken iteratively under landscapes 
where energies obey power-laws behavior, namely of disorder, 
a set of cracks is formed.
These cracks grown until the entire network fails.
For each type of disorder, the set of cracks obeys a power-law. 
For strong disorder ($\beta_{D}>10^{3}$) the fractal dimension of 
mass of cracks is $d_f=1.22$ and 
for weak disorder ($\beta_{D}\simeq 10^{-3}$) the exponent is $d_f=2$ ~\cite{Andrade2009,Oliveira2011}.
For strong disorder, this fractal dimension has been 
reported in several systems such as 
watershed line ~\cite{Fehr2009}, perimeter of the percolating
cluster at a discontinuous transition ~\cite{Fehr2012} and ranked surfaces~\cite{Schrenk2012}.

In this work, we perform a slightly the optimal path cracking (OPC) 
model proposed by Andrade \textit{et al}~\cite{Andrade2009}, we define as Local OPC.
In this model, the optimal path is determined between two nodes, 
that belong to the connected component, then the link is broken and a 
new optimal path is found. This process is repeated until a path cannot 
be found anymore.
In this work we studied the behavior of cracked
mass from source node and target node with internal distance
$l$ and its relations with the global scale $L$.
Some studies addressed the critical exponent of the backbone from 
the giant cluster~\cite{Grassberger1999,Barthelemy1999,Barthelemy2000,BARTHELEMY2003,Paul2000,Melchert2007} 
when internal distances are included but there is a lack of 
information about the behavior on recurrent failures of optimal paths.    

This paper is organized as follows. 
In Sec.~\ref{sec2} the Local OPC model is presented and its landscape properties.
In Sec.~\ref{sec3} the results obtained from Local OPC is presented 
for cracked mass and mass of optimal paths and final remarks for Sec.~\ref{sec4}.

\section{Material and Methods}\label{sec2}

Our substrate is a square lattice of linear size $L$, where
$L$ is the number of sites, with
periodic boundary conditions from top and left sides. 
Each link $i$ has energy value $\varepsilon_{i} = e^{\beta_{D}(r-1)}$
where $r$ is a random value between $[0,1)$ with uniform distribution and 
$\beta_{D}$ is a disorder parameter that has the physical meaning
of inverse temperature. 
The $\varepsilon$ values obeys 
a power-law distribution $P(\varepsilon) \simeq 1/\varepsilon$
subjected to a maximum cutoff given by $\varepsilon_{max} = e^{\beta_{D}}$.
The $\beta_{D}$ value controls the strength of disorder and broadness 
of the energy distribution.

The energy of the path is defined as the sum 
of all energies of its links. 
The optimal path is shortest path that connects the source site ($O$) 
to target site ($D$) with distance $l$ among all possible paths. 
We used the Dijkstra algorithm to find the optimal path~\cite{dijkstra1959}.

The Local OPC can be described as follows. 
Once the first optimal path between site $O$ and site $D$ is determined 
we search for the link in the optimal path having the highest energy 
and its is blocked from landscape. 
The link cannot longer be part of any path.
This is similar to impose an infinite energy to this link.
Next, the optimal path is calculated over the remaining accessible 
links of the landscape, then the link with the highest energy link is 
again removed and so on until no new paths are available.
The blocked links can be viewed as "microcracks" and this process 
continues recursively until a local failure is formed between site $O$
and site $D$ and we can no longer find any path connecting 
from source site to target site.
The cracked links forms a local failure into the system but not 
necessarily the system is offline.
Understanding local failures while maintaining network connectivity is
important in understanding many phenomena.
This model differs from the standard OPC because it inserts source site 
and the target site in the landscape.

\begin{figure}[!htb]
    \centering
    \caption{A sample Local \textit{Optimal Path Crack} (OPC) realization
    on a 128 X 128 lattice. 
    The site source and site target are in blue color.
    The LOPC was run between blue sites. 
    Each cracked link is colored in gray or red as described in sequence.
    (a) Blocked links under weak disorder conditions ($\beta=0.002$) are colored in gray.  
    (b) Blocked links under strong disorder conditions ($\beta=700$) are colored in red and
    form a backbone that can be a path with a dual-network (black line).
    (c) The broken links under strong disorder conditions are a subset of the broken link under 
    weak disorder conditions.}
    \subfigure[]{\includegraphics[width=0.23\textwidth]{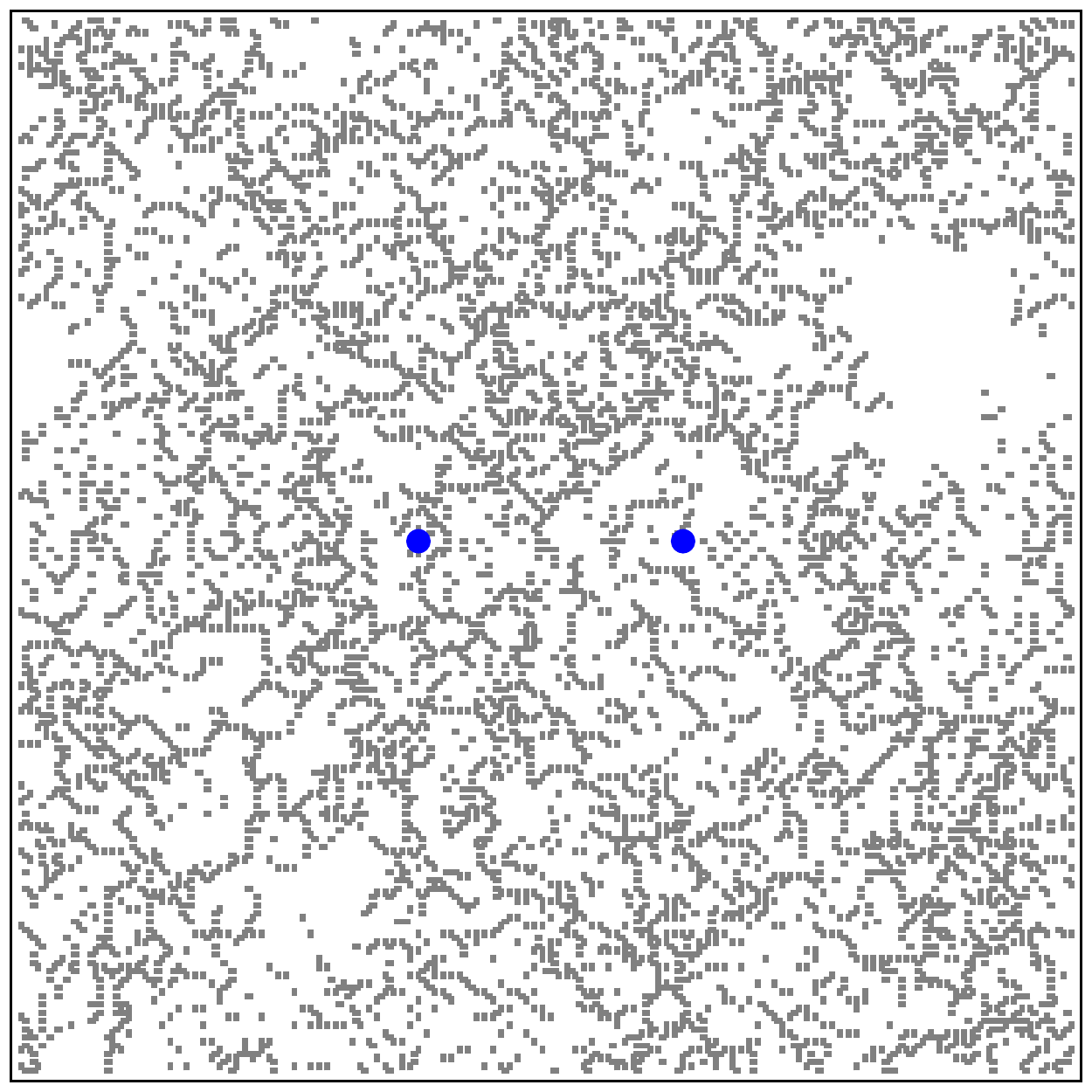}}
    \subfigure[]{\includegraphics[width=0.23\textwidth]{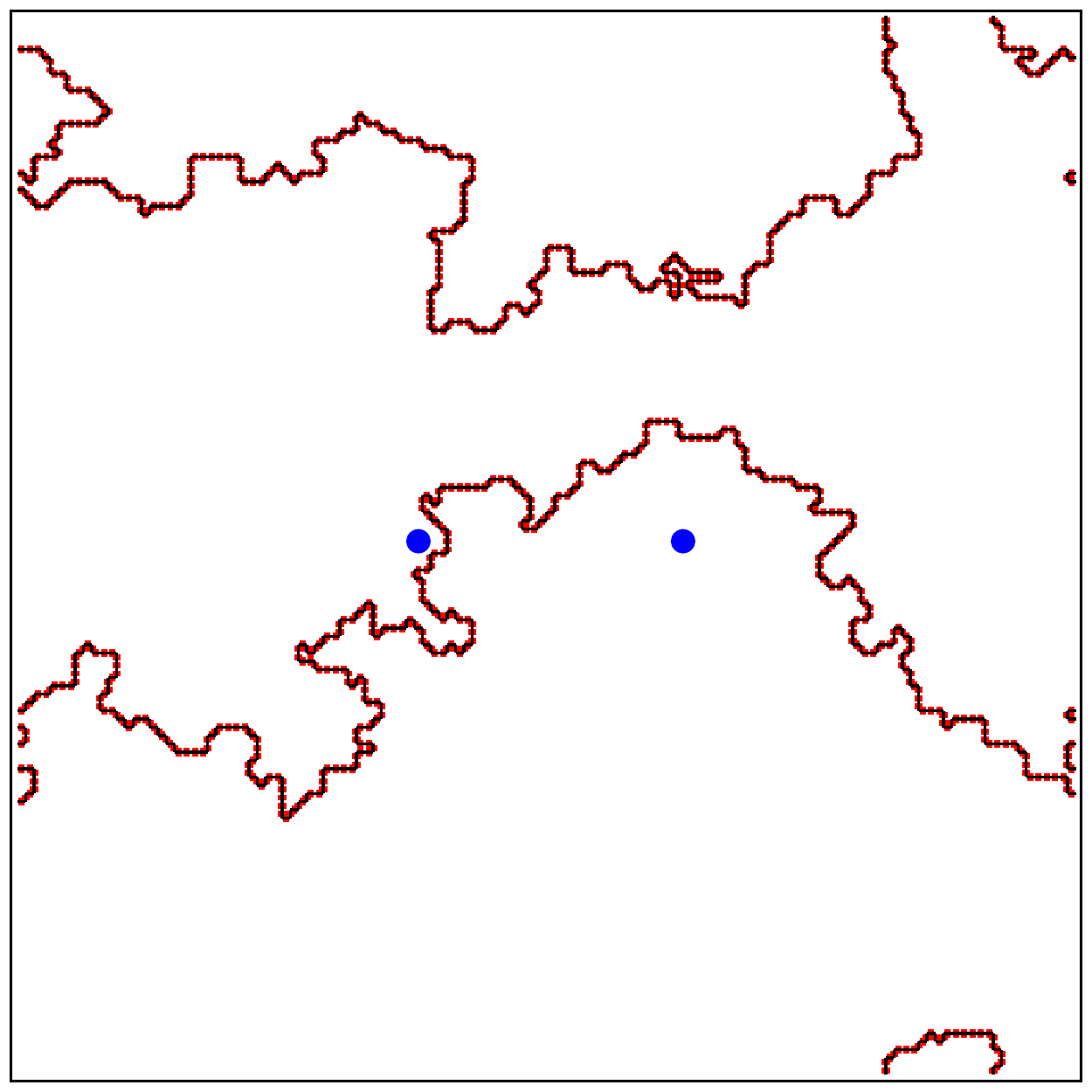}}
    \subfigure[]{\includegraphics[width=0.466\textwidth]{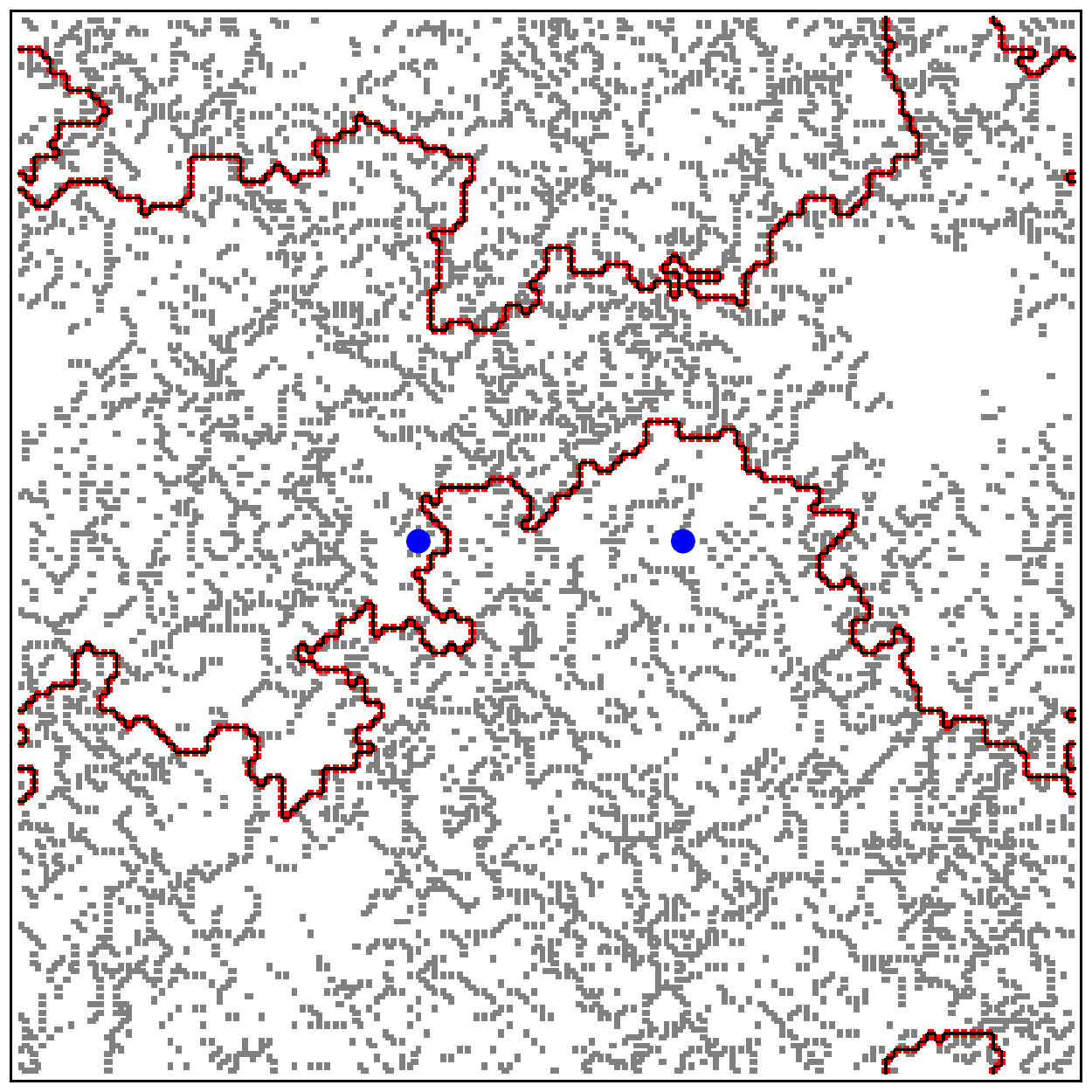}}
	\label{fig:figure1}
\end{figure} 

In Fig.~\ref{fig:figure1} we show the spatial distribution 
of blocked links in a typical 
random landscape generated under weak disorder conditions ($\beta=0.002$) - Fig.~\ref{fig:figure1}a - 
and strong disorder conditions ($\beta=700$) - Fig.~\ref{fig:figure1}b. 
The broken links from strong disorder condition are 
contained into domain broken links from weak disorder. 
As show in Fig.~\ref{fig:figure1}b, the amount of no-backbone
links reduce significantly then the disorder parameter $\beta$ increase.

The behavior shown in Fig.~\ref{fig:figure1} is related 
to the problem of minimum path in disordered landscapes~\cite{Porto1999,Oliveira2011}.
The passage from weak to strong disorder in the energy 
distribution reveals a sharp crossover between self-affine and self-similar behaviors of the optimal path. 
In the strong disorder regime the energy of the minimum path is 
controlled by the energy of a single site.
The parameter $\beta$ alone determine the limit between 
weak and strong disorder.

In order to quantify the effect of disorder from local OPC, 
we perform computer simulations for 5000 realization 
of square lattice with sizes varying in the range $16 \leq L \leq 256$  
and internal distances $1 \le l \le (L-1)$
with values of weak disorder ($\beta=0.002$)
and strong disorder ($\beta=700$). The
number of sample is 1000 for $L=256$.
The generation of random numbers that scales with a power-law 
distribution results into different landscape of energies.
Chose two sites (pair OD) with fixed distance $l$ 
we can have different configurations of broken links 
that suggest not obeying any obvious scale.

\section{Results}\label{sec3}

\begin{figure}[!htb]
    \centering
    \caption{Logarithmic dependence on distance $l$ of the mass of 
    all blocked links $M_{opc}$ for $\beta = 700$ in strong disorder 
    conditions (a) and $\beta = 0.002$ in weak disorder conditions (b)
    for several lattice sizes $L$.
    The block mass in distance $l$ near lattice size $L$ is reduced because 
    reach boundary conditions. }
    \subfigure[]{\includegraphics[width=0.5\textwidth]{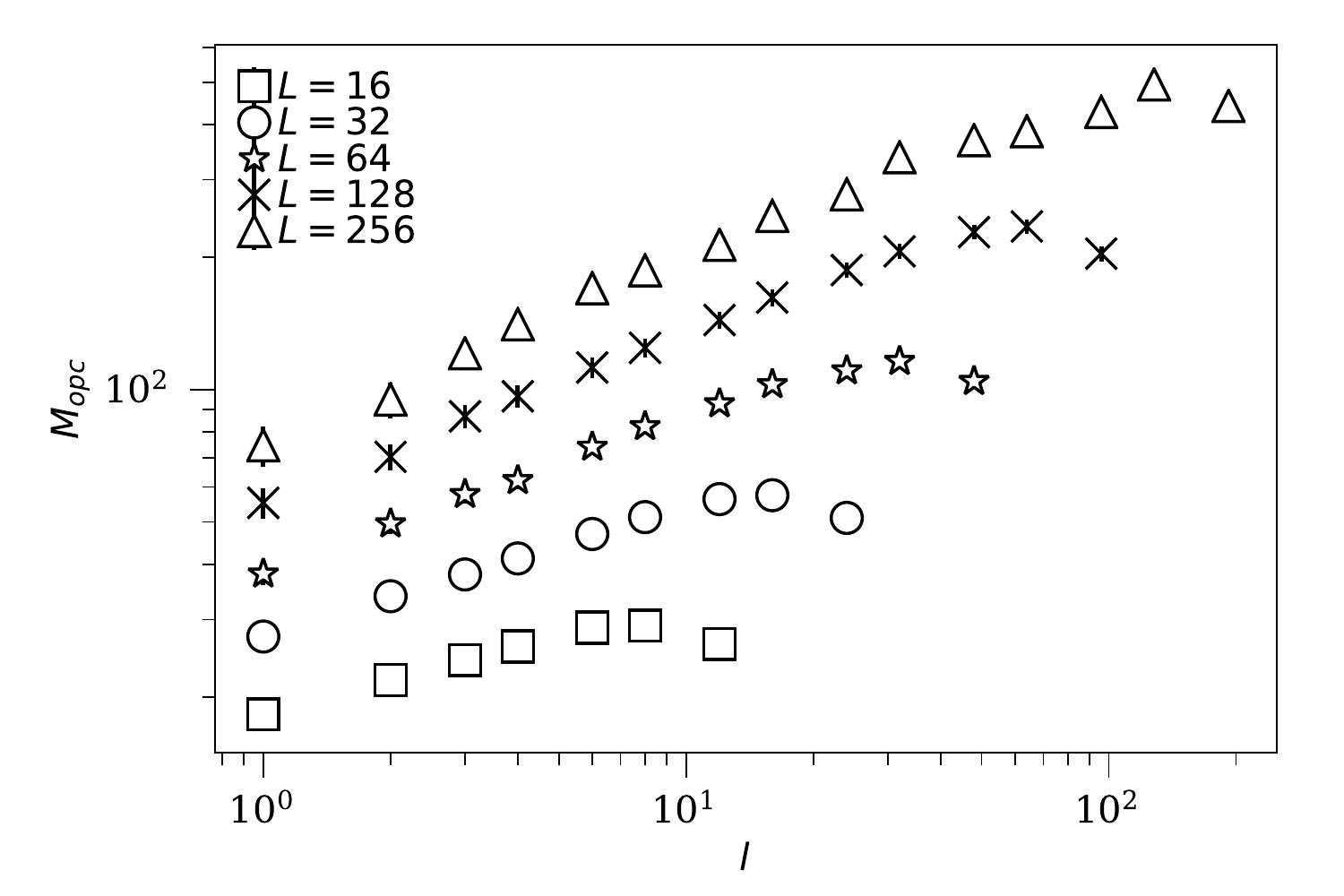}}
	\subfigure[]{\includegraphics[width=0.5\textwidth]{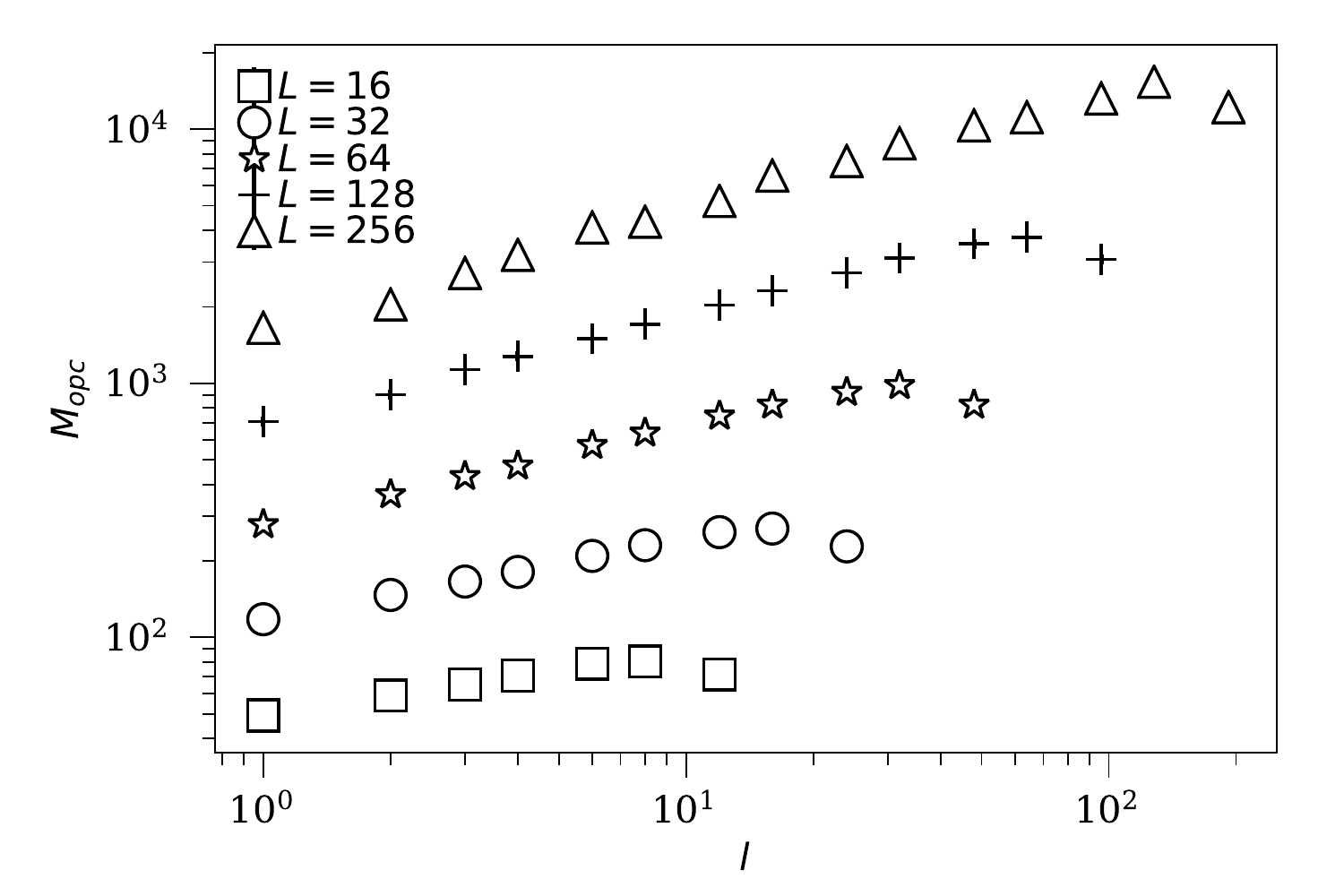}}
	
	\label{fig:figure2}
\end{figure}

The cracked mass $M_{opc}$ shows dependence 
for a disorder parameter $\beta$ and the lattice size $L$ as showed
in Fig.~\ref{fig:figure2}.
The cracked mass grows with disorder parameter decrease ~\cite{Andrade2009,Oliveira2011}
Our experiments introduce a third variable that is an internal 
distance $l$  between node $O$ and node $D$.
We see in the Fig~\ref{fig:figure2} the cracked mass enhance smoothly as 
a function from an internal distance $l$ until a critical value $l_c < L$.
The critical value $l_c$ is a boundary effect from a finite-size system. 
Initially the shape of growing follow an apparent power-law but 
the grown is quite different from standard OPC and
when the internal distance $l$ is almost the lattice size $L$
another difference is shown. 

\begin{figure}[!htb]
    \centering
    \caption{These results correspond for collapse the cracked 
    mass $M_{opc}$ and internal distance $l$. 
    The collapse of cracked mass occurs when used fractal 
    dimension $d_f=1.89$ for weak disorder conditions $\beta=0.002$ (a) 
    and linear dimension $d=1$ for strong disorder conditions (b).
    On x-axis the cracked mass collapse occurs when the internal 
    distance is normalized with lattice size $L$.
    The solid line is a least-square fit to the power-law data 
    excluding the outliers tendency.}
    \subfigure[]{\includegraphics[width=0.5\textwidth]{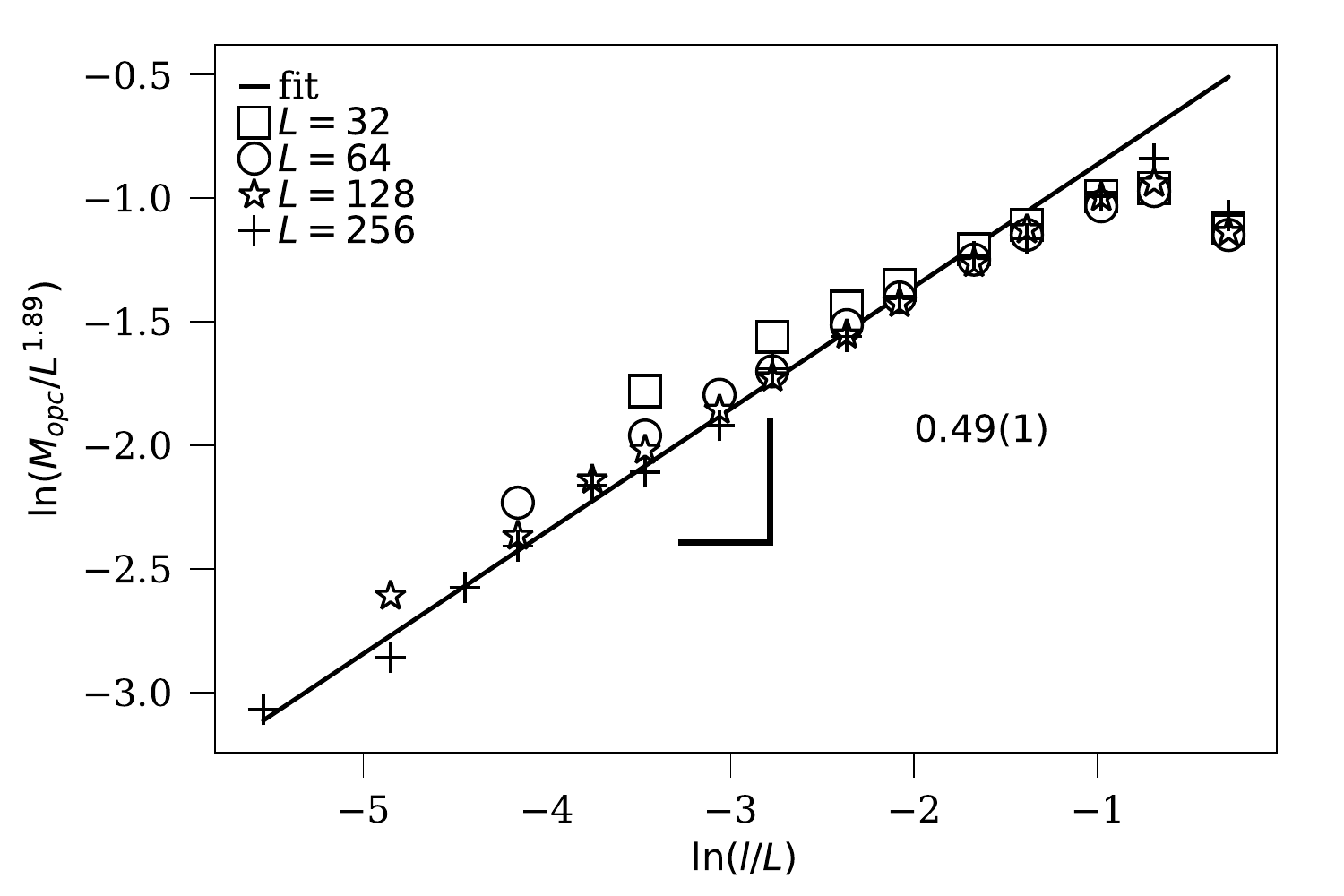}}
	\subfigure[]{\includegraphics[width=0.5\textwidth]{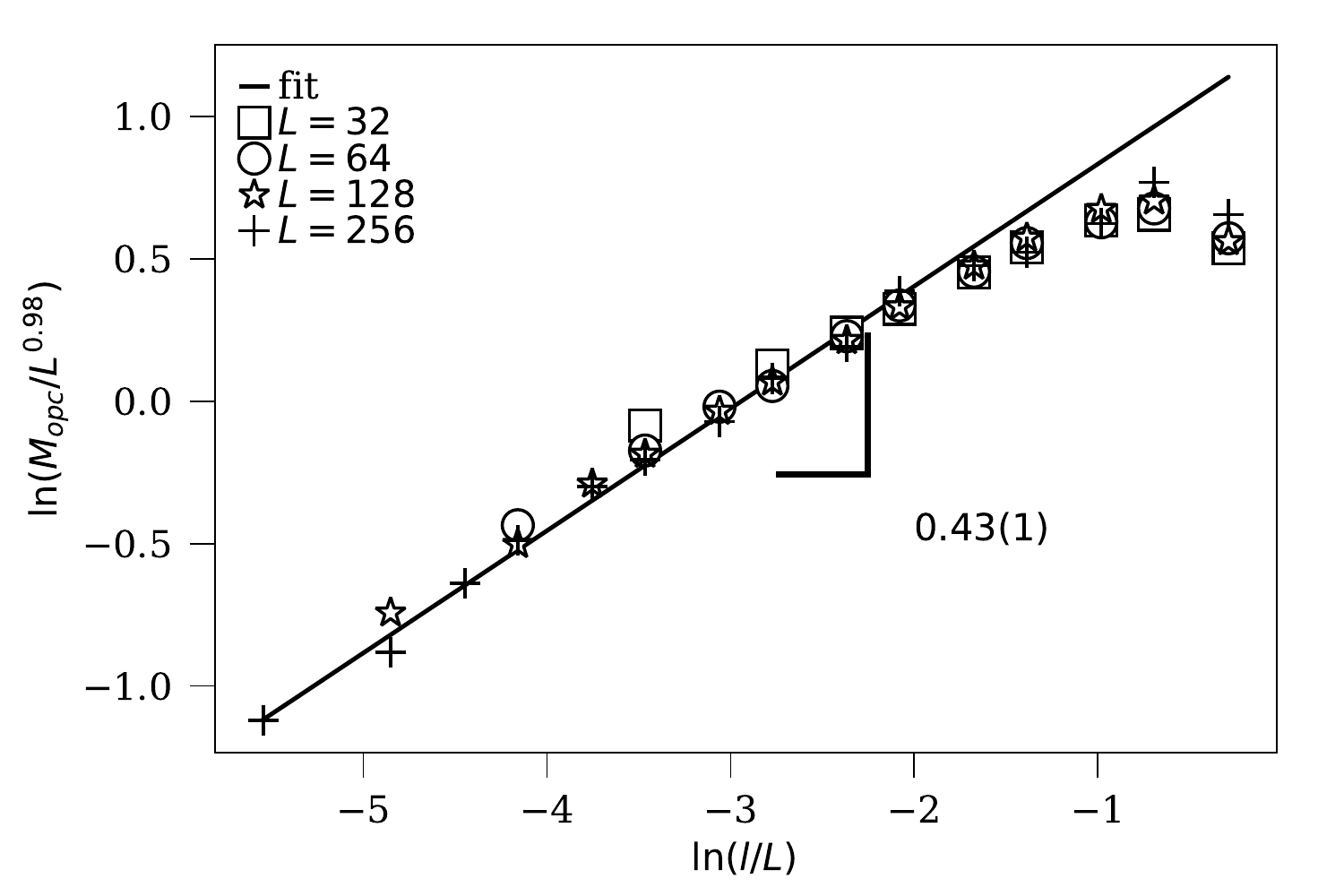}}
	
	\label{fig:figure3}
\end{figure} 

The finite size scaling method is applied from cracked mass 
and is showed in Fig~\ref{fig:figure3}.
As the site source and node target is contained in the connected 
component the maximum set of the cracked links not can bigger than
 the fractal dimension of the standard percolation $M_{cluster} \sim L^{d_{f}}$ 
 where $d_{f}=1.89$ is the fractal dimension of percolating cluster
 ~\cite{staufer1994,Schwarzer1999,Knackstedt2002,Araujo2005}.
The cracked mass in the thermodynamic limit should be
\begin{equation}
	\label{equation001}
	M_{opc} \sim L^{d_{g}}
\end{equation}
where $d_{g}$ is the global dimension of a percolating cluster that
depends from disorder parameter $\beta$.
The size scaling hypothesis is that the cracked mass should be
\begin{equation}
	\label{equation002}
	M_{opc} = L^{d_{g}} M_{0}(l/L)
\end{equation}
where
\begin{equation}
M_{0}(l/L) 
  \begin{cases}
    = K       & \quad \text{if } l \rightarrow L \\
    \sim (l/L)^{d_{i}}  & \quad \text{if } l << L
  \end{cases}
\end{equation}
and $K$ is a constant value.
When $l << L$ we have a new power-law that describe the enhancement of cracked mass  
\begin{equation}
	\label{equation1}
	M_{opc} \sim {L^{d_{g}-d_{i}}} l^{d_{i}}
\end{equation}
where $d_{g}$ is the global dimension from lattice size $L$ and 
$d_{i}$ is an internal dimension from distance $l$. 
In other cases the internal dimension can be named by co-dimension.

The finite-size scaling of the experiment data is shown in to Fig.~\ref{fig:figure3}.
The cracked mass is enhanced with the internal 
dimension $d_{i}=0.49(1)$ for weak disorder and 
$d_{i}=0.43(1)$ for strong disorder.
The data are collapsed when we use the global dimension for weak 
disorder $d_{g}=1.89$ and for strong disorder $d_{g}=0.98$.

We need to understand the physical meaning the global dimension $d_g$ 
and its relationship with disorder parameter $\beta$.
When $\beta$ is sufficiently low the sampling of the distribution 
of energy has a significant cutoff near value 1, 
in other words, the energy variance is shallow and 
the $\varepsilon$ for each link is near value 1.
The OPC removes the link with maximum energy for each path 
found, more precisely, the effect is similar the 
random cracks for each path in its interactions. 
In this situation the cracked mass increase as 
the residual mass from connected components of percolation cluster then
the mass scaling obeys $M \sim L^{d_{f}}$ where $d_{f}=1.89$
~\cite{staufer1994,Barthelemy1999,Schwarzer1999,Knackstedt2002,Araujo2005}
is the fractal dimension of percolation cluster.
For other side when $\beta$ is enough strong the 
sampling of the distribution of energy has 
full range values $(0,1]$. 
The cracked mass collapse for lattice size $L$ as showed in Fig.~\ref{fig:figure3}.
This behavior is interesting because the percolating cluster is identical into both disorder scenario
and the unique difference is a energy each link. 
Previous studies showed that the cracked mass in the standard problem 
using OPC scales with fractal dimension of $d_{f}=1.22$
~\cite{Andrade2009,Fehr2009,Oliveira2011,Fehr2012,Schrenk2012} and
the cracked mass grows linearly with lattice size $L$ with 
an internal dimension $d_{i}=1$ that it is the fractal dimension of the lower backbone  
~\cite{Grassberger1999,Melchert2007}.

The cracked mass for internal distance $l$ under weak disorder is
given by 
\begin{equation}
 \lim_{\beta \to\ 0} M_{opc} \sim 
  \begin{cases}
     L^{1.89}      & \quad \text{if } l \rightarrow L \\
     L^{1.40}l^{0.49}  & \quad \text{if } l << L
  \end{cases}
\end{equation}
 
and for strong disorder the cracked mass is given by
\begin{equation}
 \lim_{\beta \to\ 1000} M_{opc} \sim 
  \begin{cases}
     L^{0.98}      & \quad \text{if } l \rightarrow L \\
     L^{0.45}l^{0.43}  & \quad \text{if } l << L
  \end{cases}
\end{equation}
for lattice size $L$.

\begin{figure}[!htb]
    \centering
    \caption{
    These results correspond for collapse the average mass of first path 
	$M_{sp}$ and internal distance $l$. 
	The collapse of shortest mass occurs
	when used linear dimension $d_{g} = 1$ for weak disorder conditions
	$\beta = 0.002$ (a) 
	and fractal dimension $d_{g} = 1.22$ for strong disorder conditions (b). 
	On x-axis the average mass of first path collapse occurs when the
    internal distance $l$ is normalized with lattice size $L$. 
    The solid line is a least-square fit to the power-law data 
    excluding the outliers tendency.}
    \subfigure[]{\includegraphics[width=0.5\textwidth]{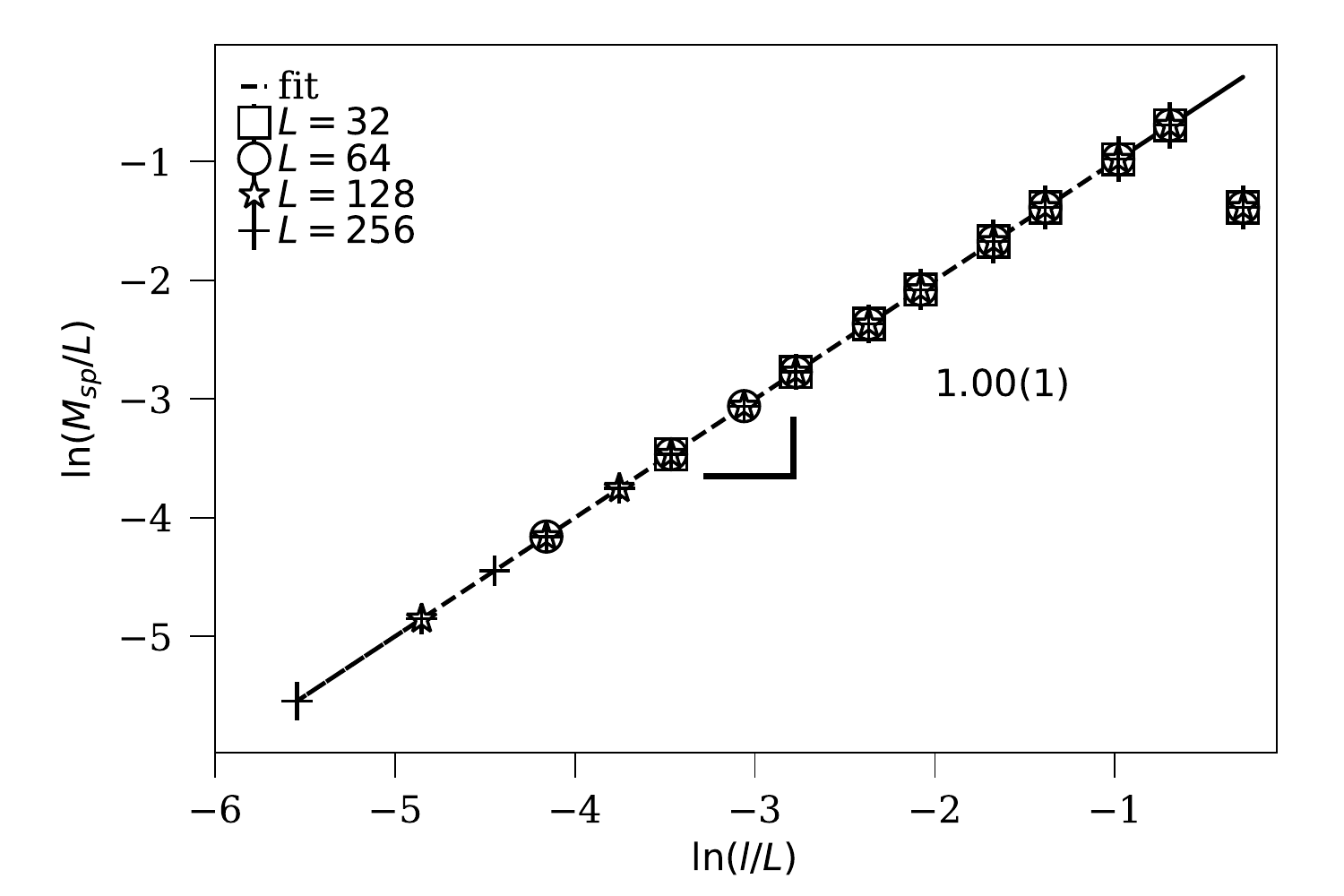}}
	\subfigure[]{\includegraphics[width=0.5\textwidth]{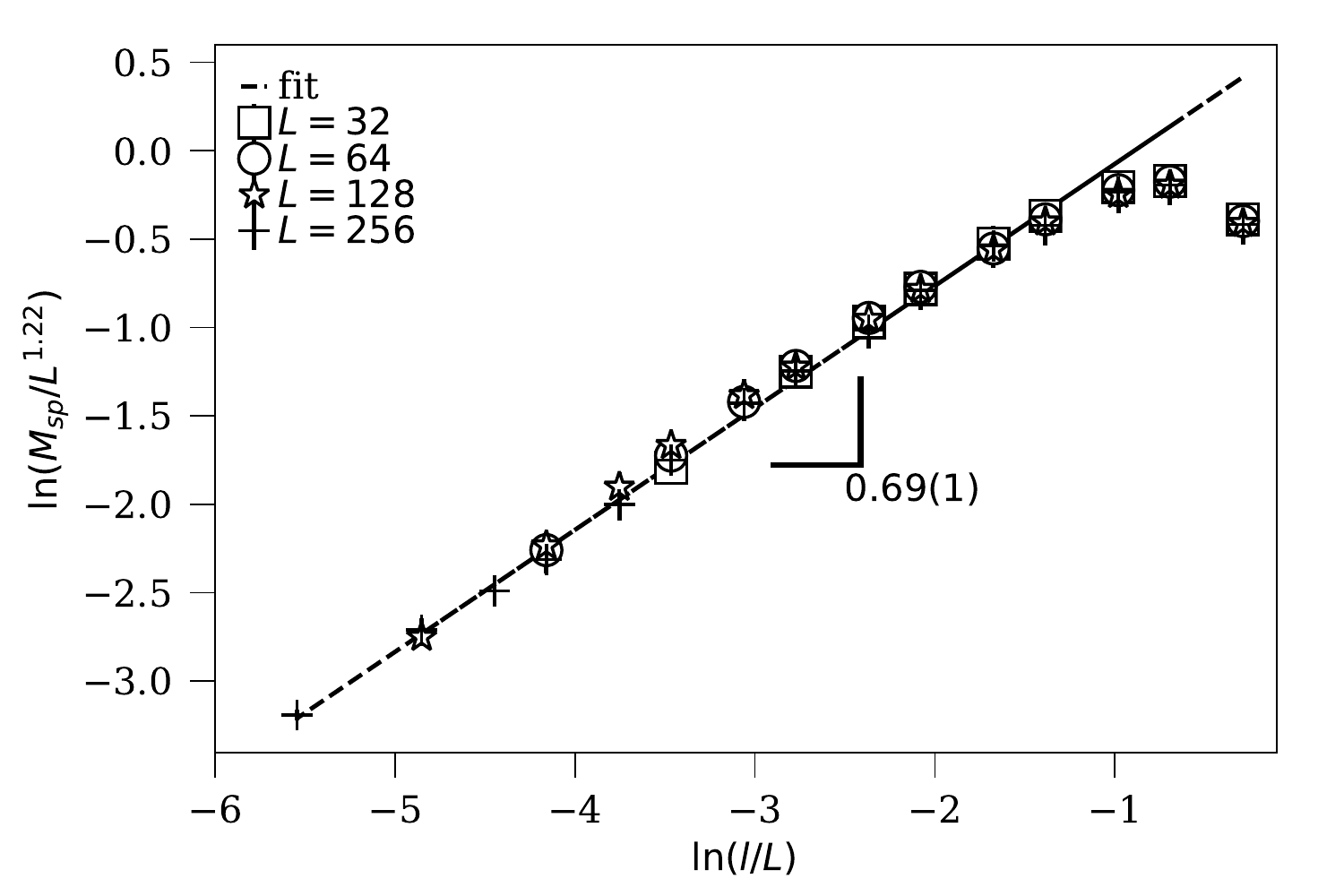}}
	
	\label{fig:figure4}
\end{figure} 

We studied the shortest path mass using the finite-size scaling.
The mass of shortest paths has a similar behavior when the average mass of 
first paths are collapsed
as showed in Fig.~\ref{fig:figure4}.
We use the equation~\ref{equation1} and the mass
of the shortest paths scales $M_{sp} \sim l$
for any lattice size $L$ for weak disorder conditions 
and the dimension is $d_g=d_{i}=1$ 
that is standard dimension for shortest paths with its energy properties
~\cite{Cieplak1994,Cieplak1995,barabasi1995,Cieplak1996,Schwartz1998}.
In strong disorder, the average mass of the shortest paths collapses using $d_g=1.22$ 
this is the standard fractal dimension ~\cite{Cieplak1994,Cieplak1995,Cieplak1996,Porto1999}.
The internal dimension found is $d_{i}=0.69(1)$~\cite{Schwartz1998,Buldyrev2006} for strong
disorder conditions. 
Thereby the mass of the first path growing to obey the power-law 
\begin{equation}
 M_{sp} \sim 
  \begin{cases}
     L^{1.22}      & \quad \text{if } l \rightarrow L \\
     L^{0.53}l^{0.69}  & \quad \text{if } l << L
  \end{cases}
\end{equation}

as previous results from the thermodynamic limit  
~\cite{Cieplak1994,Cieplak1995,Cieplak1996,Schwartz1998,Porto1999,Buldyrev2004,Buldyrev2006}. 

\begin{figure}[!htb]
    \centering
    \caption{The a sample Local \textit{Optimal Path Crack} (OPC) realization
    on a 128 X 128 lattice. The blocked links generated under weak disorder
    conditions ($\beta=0.002$) and strong disorder conditions ($\beta=700$) 
    can be categorized in two types, namely, the transversal backbone (red links), 
    the cracked links (gray) and the backbone (black). 
    The backbone is obtained by dual network from transversal backbone. 
    The source site and target site are colored with blue.
    In (a) and (b) the broken links landscape is shown. 
    This difference is due to the generation of values of the power-law distribution.}
    \subfigure[]{\includegraphics[width=0.23\textwidth]{drawL128_l32_s3_b00002000_3}}
	\subfigure[]{\includegraphics[width=0.23\textwidth]{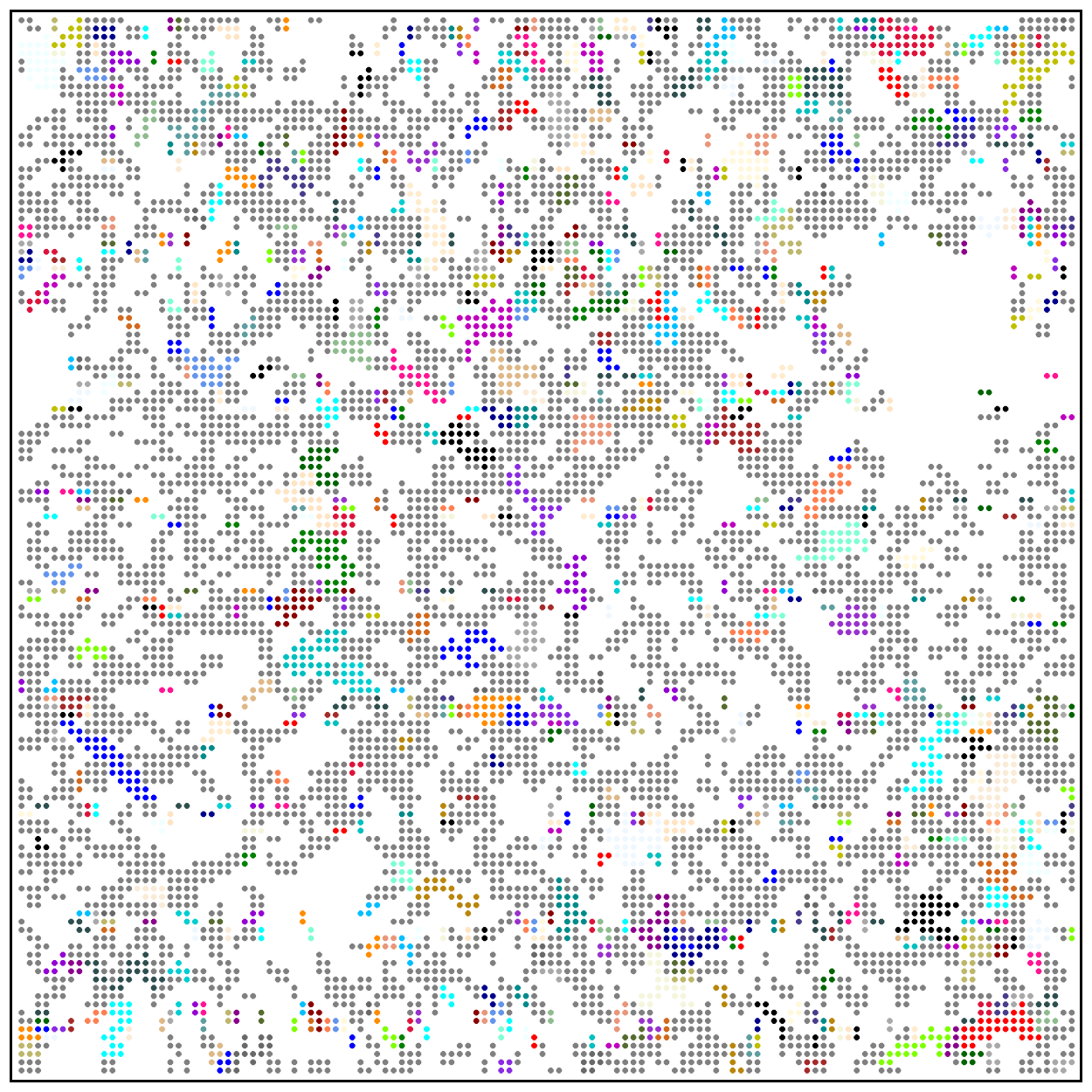}}
	\subfigure[]{\includegraphics[width=0.23\textwidth]{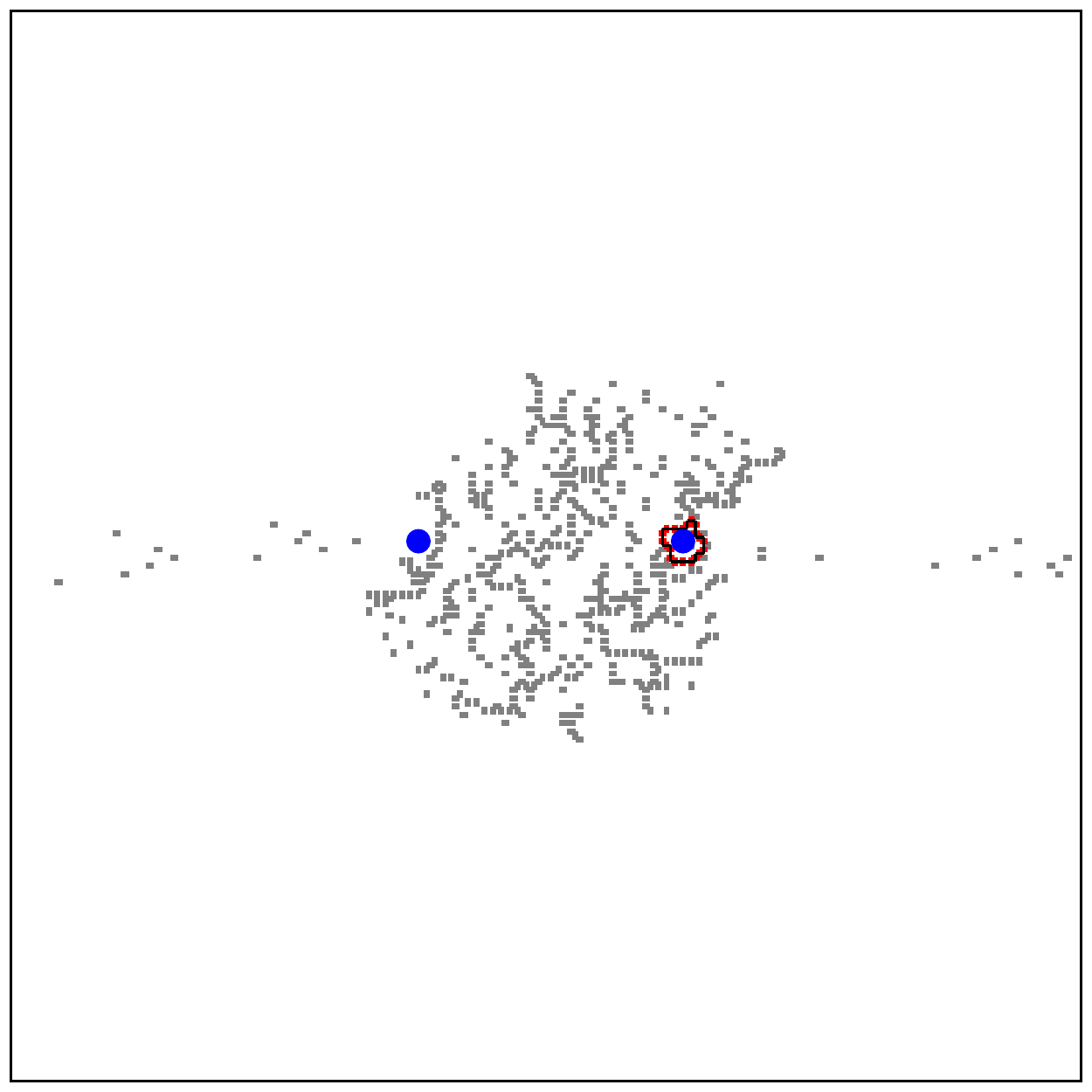}}
    \subfigure[]{\includegraphics[width=0.23\textwidth]{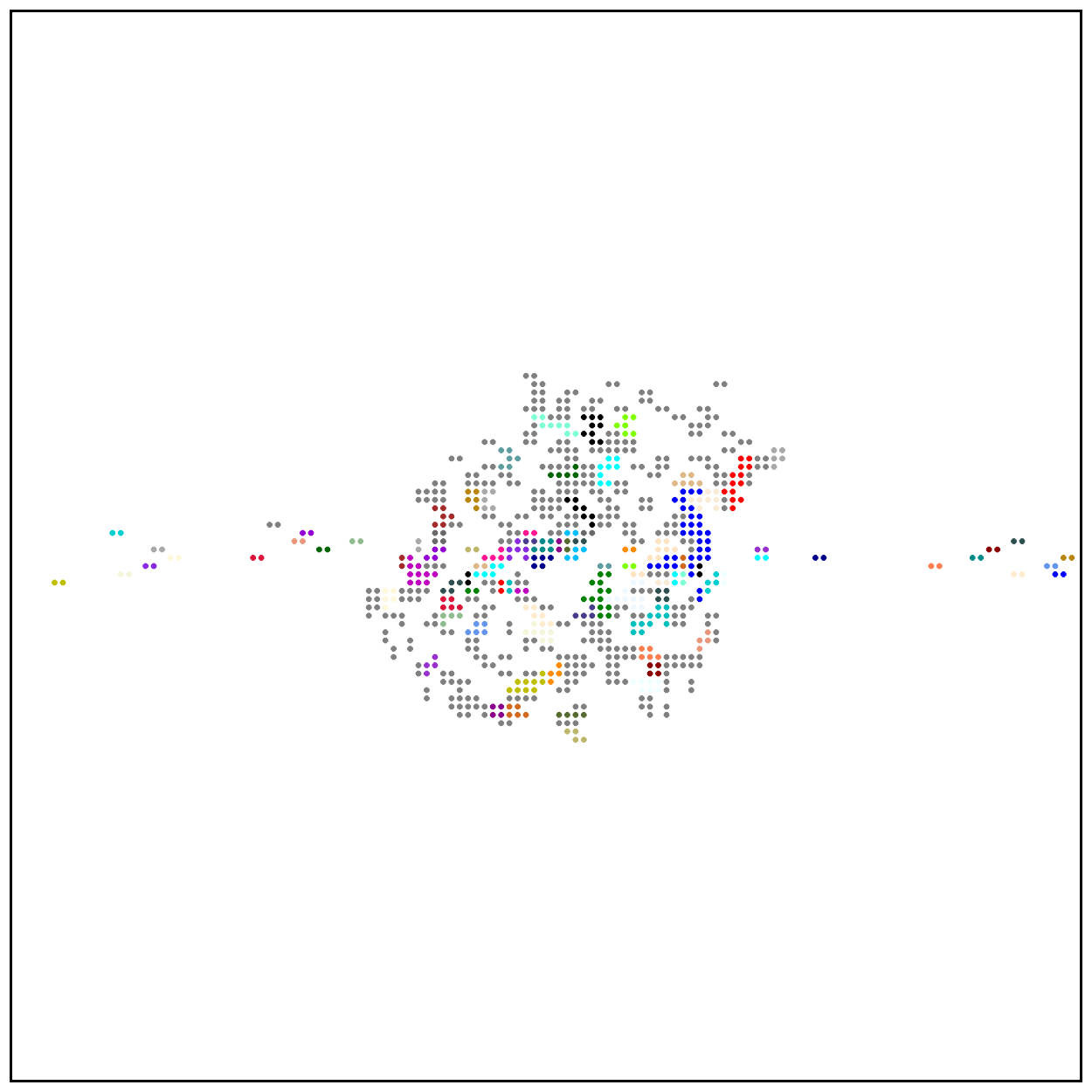}}
    \label{fig:figure5}
\end{figure}

Another set of properties found in the analysis of cracked mass 
is the sign of cluster after all interactions of the Local OPC.
In Fig.~\ref{fig:figure5} we show two extreme scenarios generated 
from optimal path cracking interactions. 
The first is completely cracked and the second the landscape 
is partially cracked but both scenarios do not have a path 
between the source site and the target site.

\begin{figure}[!htb]
    \centering
    \caption{
    The cluster size distribution $p(s)$ from several samples 
    from lattice size $L=256$. 
    The cracked links form a subset network that connected components 
    were determined. 
    The system is considered to be in the weak disorder 
    regime for this value and this range of internal distances $2 \le l \le 128$.
    The solid line are the least-squares fits the data of power-laws
    $p(s) \sim s^{-\tau}$ with $\tau = 2.67(5)$.}
	\includegraphics[width=0.5\textwidth]{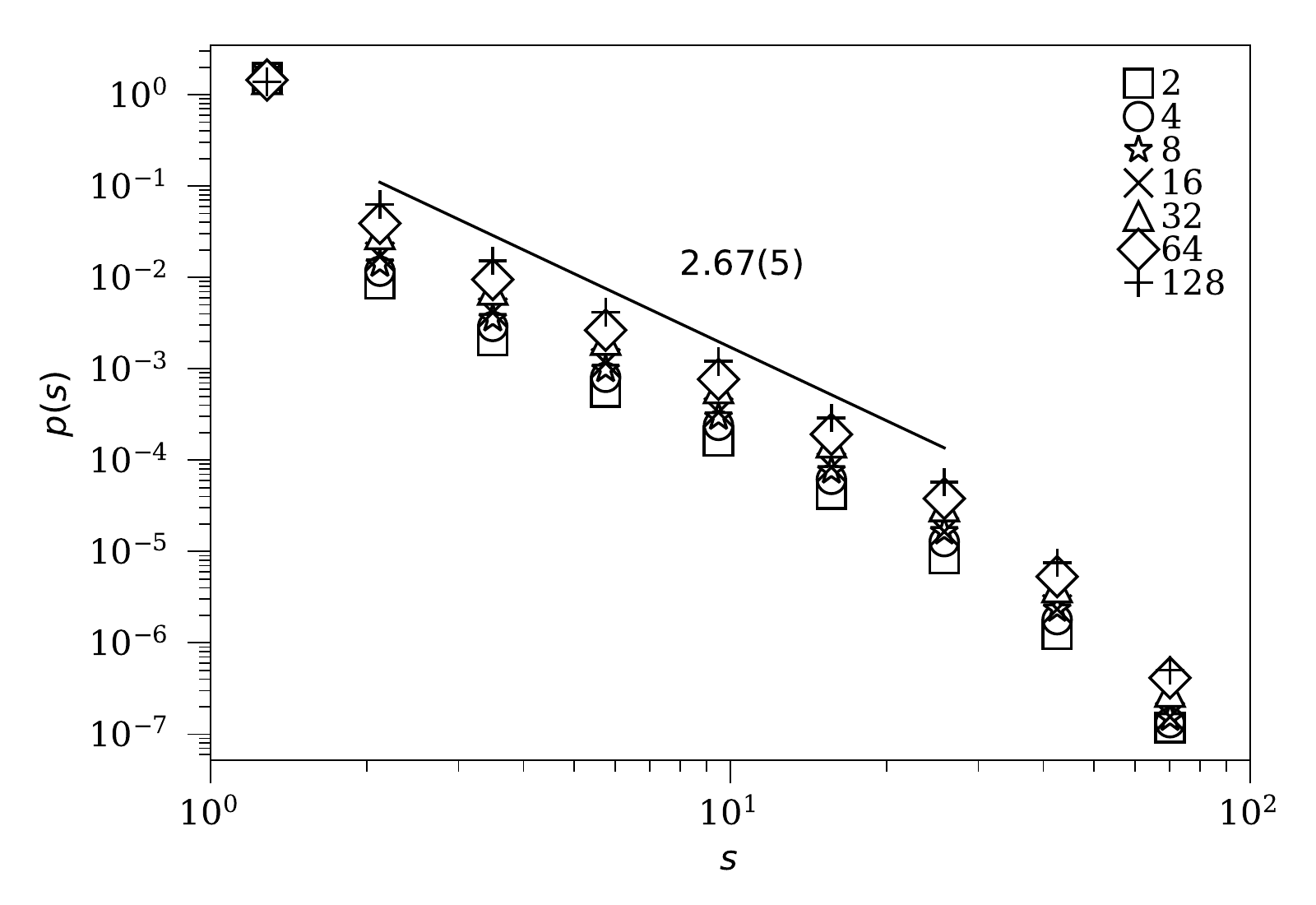}
	\label{fig:figure6}
\end{figure}

A subset network was produced from cracked links and the connected 
components were determined~\cite{Tarjan1972} for both 
scenarios of energies disorder.
The connected components made from cracked links leave a visual sign 
and the cluster size distribution shows the same inclination 
that is shown in Fig.~\ref{fig:figure6} for weak disorder conditions. 
For the network with the same lattice size $L$ the cracked mass 
for each internal distance $l$ has the same cluster size distributions.
In the strong disorder conditions, each site is a cluster 
thus the distribution is a delta function centered into one that is does not
show here.  

\section{Final Remarks}\label{sec4}

In this work we have studied in detail modification 
of the optimal-path crack (OPC) introduced by
Andrade et al.~\cite{Andrade2009} 
that we call Local Optimal-Path Crack (OPC) for
the weak and the strong disorder limits.
This modification enables us to find paths 
between two sites that are contained from the network 
from an internal distance $l$. 

We found interesting exponents for cracked mass 
links and shortest paths when using a finite-size scaling method.
The mass of cracked links obeys a power-laws that scales with 
the internal distance $l$ and lattice size $L$. 
Even with values of disorder parameters, the internal 
dimension $d_i$ has like values such $d_i=0.49(1)$ for weak disorder 
conditions and $d_i=0.43(1)$ for strong disorder conditions.   
The shortest paths obeys a power-laws that, in the limit $l<L$, 
the internal dimension $d_i$ has different values $d_i=1$ 
for weak disorder conditions and $d_i=0.69(1)$ for 
strong disorder conditions. 

Concluding, we discover that for all disorders the cracked 
mass scales with an internal dimension of 0.4 for internal distance $l$.
The role of disorder and system size can be fully cast in a 
crossover scaling law for the total number of cracked sites. 
Our model presents new challenges also from the theoretical point of 
view since the finite-size scaling breaks the dimension into two branches, 
the first, a dimension for $l$, and another dimension to $L$. 
It seems to hint towards some deeper relation. 
Potential applications of our model includes
the optimal transportation to traffic congestion under 
sequential bottlenecks and the electrical break-down of 
varistor ceramics. Finally, it could be certainly interesting 
to study the influence of the dimension of the system 
and of the substrate topology on our model by a generalization 
to three-dimensional systems or complex networks.

\section{Acknowledgments}
We acknowledge financial support from scientist leader program through FUNCAP.
We also acknowledge the Brazilian agencies CNPq, CAPES, and FUNCAP
 for financial support.

\bibliographystyle{apsrev4-1}
\bibliography{paper}

\begin{thebibliography}{48}%
\makeatletter
\providecommand \@ifxundefined [1]{%
 \@ifx{#1\undefined}
}%
\providecommand \@ifnum [1]{%
 \ifnum #1\expandafter \@firstoftwo
 \else \expandafter \@secondoftwo
 \fi
}%
\providecommand \@ifx [1]{%
 \ifx #1\expandafter \@firstoftwo
 \else \expandafter \@secondoftwo
 \fi
}%
\providecommand \natexlab [1]{#1}%
\providecommand \enquote  [1]{``#1''}%
\providecommand \bibnamefont  [1]{#1}%
\providecommand \bibfnamefont [1]{#1}%
\providecommand \citenamefont [1]{#1}%
\providecommand \href@noop [0]{\@secondoftwo}%
\providecommand \href [0]{\begingroup \@sanitize@url \@href}%
\providecommand \@href[1]{\@@startlink{#1}\@@href}%
\providecommand \@@href[1]{\endgroup#1\@@endlink}%
\providecommand \@sanitize@url [0]{\catcode `\\12\catcode `\$12\catcode
  `\&12\catcode `\#12\catcode `\^12\catcode `\_12\catcode `\%12\relax}%
\providecommand \@@startlink[1]{}%
\providecommand \@@endlink[0]{}%
\providecommand \url  [0]{\begingroup\@sanitize@url \@url }%
\providecommand \@url [1]{\endgroup\@href {#1}{\urlprefix }}%
\providecommand \urlprefix  [0]{URL }%
\providecommand \Eprint [0]{\href }%
\providecommand \doibase [0]{http://dx.doi.org/}%
\providecommand \selectlanguage [0]{\@gobble}%
\providecommand \bibinfo  [0]{\@secondoftwo}%
\providecommand \bibfield  [0]{\@secondoftwo}%
\providecommand \translation [1]{[#1]}%
\providecommand \BibitemOpen [0]{}%
\providecommand \bibitemStop [0]{}%
\providecommand \bibitemNoStop [0]{.\EOS\space}%
\providecommand \EOS [0]{\spacefactor3000\relax}%
\providecommand \BibitemShut  [1]{\csname bibitem#1\endcsname}%
\let\auto@bib@innerbib\@empty
\bibitem [{\citenamefont {Andrade}\ \emph {et~al.}(2009)\citenamefont
  {Andrade}, \citenamefont {Oliveira}, \citenamefont {Moreira},\ and\
  \citenamefont {Herrmann}}]{Andrade2009}%
  \BibitemOpen
  \bibfield  {author} {\bibinfo {author} {\bibfnamefont {J.~S.}\ \bibnamefont
  {Andrade}}, \bibinfo {author} {\bibfnamefont {E.~A.}\ \bibnamefont
  {Oliveira}}, \bibinfo {author} {\bibfnamefont {A.~A.}\ \bibnamefont
  {Moreira}}, \ and\ \bibinfo {author} {\bibfnamefont {H.~J.}\ \bibnamefont
  {Herrmann}},\ }\href@noop {} {\bibfield  {journal} {\bibinfo  {journal}
  {Phys. Rev. Lett.}\ }\textbf {\bibinfo {volume} {103}},\ \bibinfo {pages}
  {225503} (\bibinfo {year} {2009})}\BibitemShut {NoStop}%
\bibitem [{\citenamefont {Family}\ and\ \citenamefont
  {Vicsek}(1985)}]{Family1985}%
  \BibitemOpen
  \bibfield  {author} {\bibinfo {author} {\bibfnamefont {F.}~\bibnamefont
  {Family}}\ and\ \bibinfo {author} {\bibfnamefont {T.}~\bibnamefont
  {Vicsek}},\ }\href {\doibase 10.1088/0305-4470/18/2/005} {\bibfield
  {journal} {\bibinfo  {journal} {Journal of Physics A: Mathematical and
  General}\ }\textbf {\bibinfo {volume} {18}},\ \bibinfo {pages} {L75}
  (\bibinfo {year} {1985})}\BibitemShut {NoStop}%
\bibitem [{\citenamefont {Barab{\'{a}}si}\ and\ \citenamefont
  {Albert}(1999)}]{Barabasi1999}%
  \BibitemOpen
  \bibfield  {author} {\bibinfo {author} {\bibfnamefont {A.-L.}\ \bibnamefont
  {Barab{\'{a}}si}}\ and\ \bibinfo {author} {\bibfnamefont {R.}~\bibnamefont
  {Albert}},\ }\href {\doibase 10.1126/science.286.5439.509} {\bibfield
  {journal} {\bibinfo  {journal} {Science}\ }\textbf {\bibinfo {volume}
  {286}},\ \bibinfo {pages} {509} (\bibinfo {year} {1999})}\BibitemShut
  {NoStop}%
\bibitem [{\citenamefont {Barab{\'{a}}si}(2001)}]{Barabasi2001}%
  \BibitemOpen
  \bibfield  {author} {\bibinfo {author} {\bibfnamefont {A.-L.}\ \bibnamefont
  {Barab{\'{a}}si}},\ }\href {\doibase 10.1088/2058-7058/14/7/32} {\bibfield
  {journal} {\bibinfo  {journal} {Physics World}\ }\textbf {\bibinfo {volume}
  {14}},\ \bibinfo {pages} {33} (\bibinfo {year} {2001})}\BibitemShut {NoStop}%
\bibitem [{\citenamefont {Dorogovtsev}\ and\ \citenamefont
  {Mendes}(2002)}]{Dorogovtsev2002}%
  \BibitemOpen
  \bibfield  {author} {\bibinfo {author} {\bibfnamefont {S.~N.}\ \bibnamefont
  {Dorogovtsev}}\ and\ \bibinfo {author} {\bibfnamefont {J.~F.~F.}\
  \bibnamefont {Mendes}},\ }\href {\doibase 10.1080/00018730110112519}
  {\bibfield  {journal} {\bibinfo  {journal} {Advances in Physics}\ }\textbf
  {\bibinfo {volume} {51}},\ \bibinfo {pages} {1079} (\bibinfo {year}
  {2002})}\BibitemShut {NoStop}%
\bibitem [{\citenamefont {Newman}(2002)}]{Newman2002}%
  \BibitemOpen
  \bibfield  {author} {\bibinfo {author} {\bibfnamefont {M.}~\bibnamefont
  {Newman}},\ }\href {\doibase 10.1016/s0010-4655(02)00201-1} {\bibfield
  {journal} {\bibinfo  {journal} {Computer Physics Communications}\ }\textbf
  {\bibinfo {volume} {147}},\ \bibinfo {pages} {40} (\bibinfo {year}
  {2002})}\BibitemShut {NoStop}%
\bibitem [{\citenamefont {Xu}\ \emph {et~al.}(2007)\citenamefont {Xu},
  \citenamefont {Hu}, \citenamefont {Liu},\ and\ \citenamefont {Liu}}]{Xu2007}%
  \BibitemOpen
  \bibfield  {author} {\bibinfo {author} {\bibfnamefont {X.}~\bibnamefont
  {Xu}}, \bibinfo {author} {\bibfnamefont {J.}~\bibnamefont {Hu}}, \bibinfo
  {author} {\bibfnamefont {F.}~\bibnamefont {Liu}}, \ and\ \bibinfo {author}
  {\bibfnamefont {L.}~\bibnamefont {Liu}},\ }\href {\doibase
  10.1016/j.physa.2006.06.021} {\bibfield  {journal} {\bibinfo  {journal}
  {Physica A: Statistical Mechanics and its Applications}\ }\textbf {\bibinfo
  {volume} {374}},\ \bibinfo {pages} {441} (\bibinfo {year}
  {2007})}\BibitemShut {NoStop}%
\bibitem [{\citenamefont {Estrada}\ \emph {et~al.}(2012)\citenamefont
  {Estrada}, \citenamefont {Hatano},\ and\ \citenamefont
  {Benzi}}]{Estrada2012}%
  \BibitemOpen
  \bibfield  {author} {\bibinfo {author} {\bibfnamefont {E.}~\bibnamefont
  {Estrada}}, \bibinfo {author} {\bibfnamefont {N.}~\bibnamefont {Hatano}}, \
  and\ \bibinfo {author} {\bibfnamefont {M.}~\bibnamefont {Benzi}},\ }\href
  {\doibase 10.1016/j.physrep.2012.01.006} {\bibfield  {journal} {\bibinfo
  {journal} {Physics Reports}\ }\textbf {\bibinfo {volume} {514}},\ \bibinfo
  {pages} {89} (\bibinfo {year} {2012})}\BibitemShut {NoStop}%
\bibitem [{\citenamefont {Kirkpatrick}\ and\ \citenamefont
  {Toulouse}(1985)}]{Kirkpatrick1985}%
  \BibitemOpen
  \bibfield  {author} {\bibinfo {author} {\bibfnamefont {S.}~\bibnamefont
  {Kirkpatrick}}\ and\ \bibinfo {author} {\bibfnamefont {G.}~\bibnamefont
  {Toulouse}},\ }\href {\doibase 10.1051/jphys:019850046080127700} {\bibfield
  {journal} {\bibinfo  {journal} {Journal de Physique}\ }\textbf {\bibinfo
  {volume} {46}},\ \bibinfo {pages} {1277} (\bibinfo {year}
  {1985})}\BibitemShut {NoStop}%
\bibitem [{\citenamefont {Kardar}\ \emph {et~al.}(1986)\citenamefont {Kardar},
  \citenamefont {Parisi},\ and\ \citenamefont {Zhang}}]{Kardar1986}%
  \BibitemOpen
  \bibfield  {author} {\bibinfo {author} {\bibfnamefont {M.}~\bibnamefont
  {Kardar}}, \bibinfo {author} {\bibfnamefont {G.}~\bibnamefont {Parisi}}, \
  and\ \bibinfo {author} {\bibfnamefont {Y.-C.}\ \bibnamefont {Zhang}},\ }\href
  {\doibase 10.1103/physrevlett.56.889} {\bibfield  {journal} {\bibinfo
  {journal} {Physical Review Letters}\ }\textbf {\bibinfo {volume} {56}},\
  \bibinfo {pages} {889} (\bibinfo {year} {1986})}\BibitemShut {NoStop}%
\bibitem [{\citenamefont {Kardar}\ and\ \citenamefont
  {Zhang}(1987)}]{Kardar1987}%
  \BibitemOpen
  \bibfield  {author} {\bibinfo {author} {\bibfnamefont {M.}~\bibnamefont
  {Kardar}}\ and\ \bibinfo {author} {\bibfnamefont {Y.-C.}\ \bibnamefont
  {Zhang}},\ }\href {\doibase 10.1103/physrevlett.58.2087} {\bibfield
  {journal} {\bibinfo  {journal} {Physical Review Letters}\ }\textbf {\bibinfo
  {volume} {58}},\ \bibinfo {pages} {2087} (\bibinfo {year}
  {1987})}\BibitemShut {NoStop}%
\bibitem [{\citenamefont {Perlsman}\ and\ \citenamefont
  {Schwartz}(1992)}]{Perlsman1992}%
  \BibitemOpen
  \bibfield  {author} {\bibinfo {author} {\bibfnamefont {E.}~\bibnamefont
  {Perlsman}}\ and\ \bibinfo {author} {\bibfnamefont {M.}~\bibnamefont
  {Schwartz}},\ }\href {\doibase 10.1209/0295-5075/17/1/003} {\bibfield
  {journal} {\bibinfo  {journal} {Europhysics Letters ({EPL})}\ }\textbf
  {\bibinfo {volume} {17}},\ \bibinfo {pages} {11} (\bibinfo {year}
  {1992})}\BibitemShut {NoStop}%
\bibitem [{\citenamefont {Cieplak}\ \emph {et~al.}(1994)\citenamefont
  {Cieplak}, \citenamefont {Maritan},\ and\ \citenamefont
  {Banavar}}]{Cieplak1994}%
  \BibitemOpen
  \bibfield  {author} {\bibinfo {author} {\bibfnamefont {M.}~\bibnamefont
  {Cieplak}}, \bibinfo {author} {\bibfnamefont {A.}~\bibnamefont {Maritan}}, \
  and\ \bibinfo {author} {\bibfnamefont {J.~R.}\ \bibnamefont {Banavar}},\
  }\href {\doibase 10.1103/physrevlett.72.2320} {\bibfield  {journal} {\bibinfo
   {journal} {Physical Review Letters}\ }\textbf {\bibinfo {volume} {72}},\
  \bibinfo {pages} {2320} (\bibinfo {year} {1994})}\BibitemShut {NoStop}%
\bibitem [{\citenamefont {Cieplak}\ \emph {et~al.}(1995)\citenamefont
  {Cieplak}, \citenamefont {Maritan}, \citenamefont {Swift}, \citenamefont
  {Bhattacharya}, \citenamefont {Stella},\ and\ \citenamefont
  {Banavar}}]{Cieplak1995}%
  \BibitemOpen
  \bibfield  {author} {\bibinfo {author} {\bibfnamefont {M.}~\bibnamefont
  {Cieplak}}, \bibinfo {author} {\bibfnamefont {A.}~\bibnamefont {Maritan}},
  \bibinfo {author} {\bibfnamefont {M.~R.}\ \bibnamefont {Swift}}, \bibinfo
  {author} {\bibfnamefont {A.}~\bibnamefont {Bhattacharya}}, \bibinfo {author}
  {\bibfnamefont {A.~L.}\ \bibnamefont {Stella}}, \ and\ \bibinfo {author}
  {\bibfnamefont {J.~R.}\ \bibnamefont {Banavar}},\ }\href
  {http://stacks.iop.org/0305-4470/28/i=20/a=003} {\bibfield  {journal}
  {\bibinfo  {journal} {Journal of Physics A: Mathematical and General}\
  }\textbf {\bibinfo {volume} {28}},\ \bibinfo {pages} {5693} (\bibinfo {year}
  {1995})}\BibitemShut {NoStop}%
\bibitem [{\citenamefont {Cieplak}\ \emph {et~al.}(1996)\citenamefont
  {Cieplak}, \citenamefont {Maritan},\ and\ \citenamefont
  {Banavar}}]{Cieplak1996}%
  \BibitemOpen
  \bibfield  {author} {\bibinfo {author} {\bibfnamefont {M.}~\bibnamefont
  {Cieplak}}, \bibinfo {author} {\bibfnamefont {A.}~\bibnamefont {Maritan}}, \
  and\ \bibinfo {author} {\bibfnamefont {J.~R.}\ \bibnamefont {Banavar}},\
  }\href {\doibase 10.1103/physrevlett.76.3754} {\bibfield  {journal} {\bibinfo
   {journal} {Physical Review Letters}\ }\textbf {\bibinfo {volume} {76}},\
  \bibinfo {pages} {3754} (\bibinfo {year} {1996})}\BibitemShut {NoStop}%
\bibitem [{\citenamefont {Schwartz}\ \emph {et~al.}(1998)\citenamefont
  {Schwartz}, \citenamefont {Nazaryev},\ and\ \citenamefont
  {Havlin}}]{Schwartz1998}%
  \BibitemOpen
  \bibfield  {author} {\bibinfo {author} {\bibfnamefont {N.}~\bibnamefont
  {Schwartz}}, \bibinfo {author} {\bibfnamefont {A.~L.}\ \bibnamefont
  {Nazaryev}}, \ and\ \bibinfo {author} {\bibfnamefont {S.}~\bibnamefont
  {Havlin}},\ }\href {\doibase 10.1103/physreve.58.7642} {\bibfield  {journal}
  {\bibinfo  {journal} {Physical Review E}\ }\textbf {\bibinfo {volume} {58}},\
  \bibinfo {pages} {7642} (\bibinfo {year} {1998})}\BibitemShut {NoStop}%
\bibitem [{\citenamefont {Porto}\ \emph {et~al.}(1999)\citenamefont {Porto},
  \citenamefont {Schwartz}, \citenamefont {Havlin},\ and\ \citenamefont
  {Bunde}}]{Porto1999}%
  \BibitemOpen
  \bibfield  {author} {\bibinfo {author} {\bibfnamefont {M.}~\bibnamefont
  {Porto}}, \bibinfo {author} {\bibfnamefont {N.}~\bibnamefont {Schwartz}},
  \bibinfo {author} {\bibfnamefont {S.}~\bibnamefont {Havlin}}, \ and\ \bibinfo
  {author} {\bibfnamefont {A.}~\bibnamefont {Bunde}},\ }\href@noop {}
  {\bibfield  {journal} {\bibinfo  {journal} {Phys. Rev. E}\ }\textbf {\bibinfo
  {volume} {60}},\ \bibinfo {pages} {R2448} (\bibinfo {year}
  {1999})}\BibitemShut {NoStop}%
\bibitem [{\citenamefont {Buldyrev}\ \emph {et~al.}(2004)\citenamefont
  {Buldyrev}, \citenamefont {Havlin}, \citenamefont {L\'opez},\ and\
  \citenamefont {Stanley}}]{Buldyrev2004}%
  \BibitemOpen
  \bibfield  {author} {\bibinfo {author} {\bibfnamefont {S.~V.}\ \bibnamefont
  {Buldyrev}}, \bibinfo {author} {\bibfnamefont {S.}~\bibnamefont {Havlin}},
  \bibinfo {author} {\bibfnamefont {E.}~\bibnamefont {L\'opez}}, \ and\
  \bibinfo {author} {\bibfnamefont {H.~E.}\ \bibnamefont {Stanley}},\ }\href
  {\doibase 10.1103/PhysRevE.70.035102} {\bibfield  {journal} {\bibinfo
  {journal} {Phys. Rev. E}\ }\textbf {\bibinfo {volume} {70}},\ \bibinfo
  {pages} {035102} (\bibinfo {year} {2004})}\BibitemShut {NoStop}%
\bibitem [{\citenamefont {Buldyrev}\ \emph {et~al.}(2006)\citenamefont
  {Buldyrev}, \citenamefont {Havlin},\ and\ \citenamefont
  {Stanley}}]{Buldyrev2006}%
  \BibitemOpen
  \bibfield  {author} {\bibinfo {author} {\bibfnamefont {S.~V.}\ \bibnamefont
  {Buldyrev}}, \bibinfo {author} {\bibfnamefont {S.}~\bibnamefont {Havlin}}, \
  and\ \bibinfo {author} {\bibfnamefont {H.~E.}\ \bibnamefont {Stanley}},\
  }\href {\doibase 10.1103/physreve.73.036128} {\bibfield  {journal} {\bibinfo
  {journal} {Physical Review E}\ }\textbf {\bibinfo {volume} {73}} (\bibinfo
  {year} {2006}),\ 10.1103/physreve.73.036128}\BibitemShut {NoStop}%
\bibitem [{\citenamefont {Oliveira}\ \emph {et~al.}(2011)\citenamefont
  {Oliveira}, \citenamefont {Schrenk}, \citenamefont {Araujo}, \citenamefont
  {Herrmann},\ and\ \citenamefont {Andrade}}]{Oliveira2011}%
  \BibitemOpen
  \bibfield  {author} {\bibinfo {author} {\bibfnamefont {E.~A.}\ \bibnamefont
  {Oliveira}}, \bibinfo {author} {\bibfnamefont {K.~J.}\ \bibnamefont
  {Schrenk}}, \bibinfo {author} {\bibfnamefont {N.~A.~M.}\ \bibnamefont
  {Araujo}}, \bibinfo {author} {\bibfnamefont {H.~J.}\ \bibnamefont
  {Herrmann}}, \ and\ \bibinfo {author} {\bibfnamefont {J.~S.}\ \bibnamefont
  {Andrade}},\ }\href@noop {} {\bibfield  {journal} {\bibinfo  {journal} {Phys.
  Rev. E}\ }\textbf {\bibinfo {volume} {83}},\ \bibinfo {pages} {046113}
  (\bibinfo {year} {2011})}\BibitemShut {NoStop}%
\bibitem [{\citenamefont {Zheng}\ \emph {et~al.}(2007)\citenamefont {Zheng},
  \citenamefont {Gao},\ and\ \citenamefont {Zhao}}]{ZHENG2007700}%
  \BibitemOpen
  \bibfield  {author} {\bibinfo {author} {\bibfnamefont {J.-F.}\ \bibnamefont
  {Zheng}}, \bibinfo {author} {\bibfnamefont {Z.-Y.}\ \bibnamefont {Gao}}, \
  and\ \bibinfo {author} {\bibfnamefont {X.-M.}\ \bibnamefont {Zhao}},\ }\href
  {\doibase https://doi.org/10.1016/j.physa.2007.07.031} {\bibfield  {journal}
  {\bibinfo  {journal} {Physica A: Statistical Mechanics and its Applications}\
  }\textbf {\bibinfo {volume} {385}},\ \bibinfo {pages} {700 } (\bibinfo {year}
  {2007})}\BibitemShut {NoStop}%
\bibitem [{\citenamefont {Rohden}\ \emph {et~al.}(2016)\citenamefont {Rohden},
  \citenamefont {Jung}, \citenamefont {Tamrakar},\ and\ \citenamefont
  {Kettemann}}]{Martin2016}%
  \BibitemOpen
  \bibfield  {author} {\bibinfo {author} {\bibfnamefont {M.}~\bibnamefont
  {Rohden}}, \bibinfo {author} {\bibfnamefont {D.}~\bibnamefont {Jung}},
  \bibinfo {author} {\bibfnamefont {S.}~\bibnamefont {Tamrakar}}, \ and\
  \bibinfo {author} {\bibfnamefont {S.}~\bibnamefont {Kettemann}},\ }\href
  {\doibase 10.1103/PhysRevE.94.032209} {\bibfield  {journal} {\bibinfo
  {journal} {Phys. Rev. E}\ }\textbf {\bibinfo {volume} {94}},\ \bibinfo
  {pages} {032209} (\bibinfo {year} {2016})}\BibitemShut {NoStop}%
\bibitem [{\citenamefont {Stanley}\ \emph {et~al.}(1999)\citenamefont
  {Stanley}, \citenamefont {Andrade}, \citenamefont {Havlin}, \citenamefont
  {Makse},\ and\ \citenamefont {Suki}}]{Stanley1999}%
  \BibitemOpen
  \bibfield  {author} {\bibinfo {author} {\bibfnamefont {H.}~\bibnamefont
  {Stanley}}, \bibinfo {author} {\bibfnamefont {J.~S.}\ \bibnamefont
  {Andrade}}, \bibinfo {author} {\bibfnamefont {S.}~\bibnamefont {Havlin}},
  \bibinfo {author} {\bibfnamefont {H.~A.}\ \bibnamefont {Makse}}, \ and\
  \bibinfo {author} {\bibfnamefont {B.}~\bibnamefont {Suki}},\ }\href {\doibase
  10.1016/s0378-4371(99)00029-1} {\bibfield  {journal} {\bibinfo  {journal}
  {Physica A: Statistical Mechanics and its Applications}\ }\textbf {\bibinfo
  {volume} {266}},\ \bibinfo {pages} {5} (\bibinfo {year} {1999})}\BibitemShut
  {NoStop}%
\bibitem [{\citenamefont {Albert}\ \emph {et~al.}(2000)\citenamefont {Albert},
  \citenamefont {Jeong},\ and\ \citenamefont {Barabási}}]{Albert2000}%
  \BibitemOpen
  \bibfield  {author} {\bibinfo {author} {\bibfnamefont {R.}~\bibnamefont
  {Albert}}, \bibinfo {author} {\bibfnamefont {H.}~\bibnamefont {Jeong}}, \
  and\ \bibinfo {author} {\bibfnamefont {A.-L.}\ \bibnamefont {Barabási}},\
  }\href {https://doi.org/10.1038/35019019} {\bibfield  {journal} {\bibinfo
  {journal} {Nature}\ }\textbf {\bibinfo {volume} {406}},\ \bibinfo {pages}
  {378} (\bibinfo {year} {2000})}\BibitemShut {NoStop}%
\bibitem [{\citenamefont {Holme}\ \emph {et~al.}(2002)\citenamefont {Holme},
  \citenamefont {Kim}, \citenamefont {Yoon},\ and\ \citenamefont
  {Han}}]{Holme2002}%
  \BibitemOpen
  \bibfield  {author} {\bibinfo {author} {\bibfnamefont {P.}~\bibnamefont
  {Holme}}, \bibinfo {author} {\bibfnamefont {B.~J.}\ \bibnamefont {Kim}},
  \bibinfo {author} {\bibfnamefont {C.~N.}\ \bibnamefont {Yoon}}, \ and\
  \bibinfo {author} {\bibfnamefont {S.~K.}\ \bibnamefont {Han}},\ }\href
  {\doibase 10.1103/physreve.65.056109} {\bibfield  {journal} {\bibinfo
  {journal} {Physical Review E}\ }\textbf {\bibinfo {volume} {65}} (\bibinfo
  {year} {2002}),\ 10.1103/physreve.65.056109}\BibitemShut {NoStop}%
\bibitem [{\citenamefont {Moreno}\ \emph {et~al.}(2003)\citenamefont {Moreno},
  \citenamefont {Pastor-Satorras}, \citenamefont {V{\'{a}}zquez},\ and\
  \citenamefont {Vespignani}}]{Moreno2003}%
  \BibitemOpen
  \bibfield  {author} {\bibinfo {author} {\bibfnamefont {Y.}~\bibnamefont
  {Moreno}}, \bibinfo {author} {\bibfnamefont {R.}~\bibnamefont
  {Pastor-Satorras}}, \bibinfo {author} {\bibfnamefont {A.}~\bibnamefont
  {V{\'{a}}zquez}}, \ and\ \bibinfo {author} {\bibfnamefont {A.}~\bibnamefont
  {Vespignani}},\ }\href {\doibase 10.1209/epl/i2003-00140-7} {\bibfield
  {journal} {\bibinfo  {journal} {Europhysics Letters ({EPL})}\ }\textbf
  {\bibinfo {volume} {62}},\ \bibinfo {pages} {292} (\bibinfo {year}
  {2003})}\BibitemShut {NoStop}%
\bibitem [{\citenamefont {Ellinas}\ \emph {et~al.}(2015)\citenamefont
  {Ellinas}, \citenamefont {Allan}, \citenamefont {Durugbo},\ and\
  \citenamefont {Johansson}}]{Ellinas2015}%
  \BibitemOpen
  \bibfield  {author} {\bibinfo {author} {\bibfnamefont {C.}~\bibnamefont
  {Ellinas}}, \bibinfo {author} {\bibfnamefont {N.}~\bibnamefont {Allan}},
  \bibinfo {author} {\bibfnamefont {C.}~\bibnamefont {Durugbo}}, \ and\
  \bibinfo {author} {\bibfnamefont {A.}~\bibnamefont {Johansson}},\ }\href
  {\doibase 10.1371/journal.pone.0142469} {\bibfield  {journal} {\bibinfo
  {journal} {{PLOS} {ONE}}\ }\textbf {\bibinfo {volume} {10}},\ \bibinfo
  {pages} {e0142469} (\bibinfo {year} {2015})}\BibitemShut {NoStop}%
\bibitem [{\citenamefont {Paolo~Masucci}\ and\ \citenamefont
  {Molinero}(2016)}]{PaoloMasucci2016}%
  \BibitemOpen
  \bibfield  {author} {\bibinfo {author} {\bibfnamefont {A.}~\bibnamefont
  {Paolo~Masucci}}\ and\ \bibinfo {author} {\bibfnamefont {C.}~\bibnamefont
  {Molinero}},\ }\href {\doibase 10.1140/epjb/e2016-60431-2} {\bibfield
  {journal} {\bibinfo  {journal} {The European Physical Journal B}\ }\textbf
  {\bibinfo {volume} {89}},\ \bibinfo {pages} {1} (\bibinfo {year}
  {2016})}\BibitemShut {NoStop}%
\bibitem [{\citenamefont {Parshani}\ \emph {et~al.}(2010)\citenamefont
  {Parshani}, \citenamefont {Buldyrev},\ and\ \citenamefont
  {Havlin}}]{Parshani2010b}%
  \BibitemOpen
  \bibfield  {author} {\bibinfo {author} {\bibfnamefont {R.}~\bibnamefont
  {Parshani}}, \bibinfo {author} {\bibfnamefont {S.~V.}\ \bibnamefont
  {Buldyrev}}, \ and\ \bibinfo {author} {\bibfnamefont {S.}~\bibnamefont
  {Havlin}},\ }\href {\doibase 10.1103/physrevlett.105.048701} {\bibfield
  {journal} {\bibinfo  {journal} {Physical Review Letters}\ }\textbf {\bibinfo
  {volume} {105}} (\bibinfo {year} {2010}),\
  10.1103/physrevlett.105.048701}\BibitemShut {NoStop}%
\bibitem [{\citenamefont {Buldyrev}\ \emph {et~al.}(2010)\citenamefont
  {Buldyrev}, \citenamefont {Parshani}, \citenamefont {Paul}, \citenamefont
  {Stanley},\ and\ \citenamefont {Havlin}}]{Buldyrev2010a}%
  \BibitemOpen
  \bibfield  {author} {\bibinfo {author} {\bibfnamefont {S.~V.}\ \bibnamefont
  {Buldyrev}}, \bibinfo {author} {\bibfnamefont {R.}~\bibnamefont {Parshani}},
  \bibinfo {author} {\bibfnamefont {G.}~\bibnamefont {Paul}}, \bibinfo {author}
  {\bibfnamefont {H.~E.}\ \bibnamefont {Stanley}}, \ and\ \bibinfo {author}
  {\bibfnamefont {S.}~\bibnamefont {Havlin}},\ }\href {\doibase
  10.1038/nature08932} {\bibfield  {journal} {\bibinfo  {journal} {Nature}\
  }\textbf {\bibinfo {volume} {464}},\ \bibinfo {pages} {1025} (\bibinfo {year}
  {2010})}\BibitemShut {NoStop}%
\bibitem [{\citenamefont {Gao}\ \emph {et~al.}(2011)\citenamefont {Gao},
  \citenamefont {Buldyrev}, \citenamefont {Havlin},\ and\ \citenamefont
  {Stanley}}]{Gao2011}%
  \BibitemOpen
  \bibfield  {author} {\bibinfo {author} {\bibfnamefont {J.}~\bibnamefont
  {Gao}}, \bibinfo {author} {\bibfnamefont {S.~V.}\ \bibnamefont {Buldyrev}},
  \bibinfo {author} {\bibfnamefont {S.}~\bibnamefont {Havlin}}, \ and\ \bibinfo
  {author} {\bibfnamefont {H.~E.}\ \bibnamefont {Stanley}},\ }\href {\doibase
  10.1103/physrevlett.107.195701} {\bibfield  {journal} {\bibinfo  {journal}
  {Physical Review Letters}\ }\textbf {\bibinfo {volume} {107}} (\bibinfo
  {year} {2011}),\ 10.1103/physrevlett.107.195701}\BibitemShut {NoStop}%
\bibitem [{\citenamefont {Carmona}\ \emph {et~al.}(2020)\citenamefont
  {Carmona}, \citenamefont {de~Noronha}, \citenamefont {Moreira}, \citenamefont
  {Ara{\'{u}}jo},\ and\ \citenamefont {Andrade}}]{Carmona2020}%
  \BibitemOpen
  \bibfield  {author} {\bibinfo {author} {\bibfnamefont {H.~A.}\ \bibnamefont
  {Carmona}}, \bibinfo {author} {\bibfnamefont {A.~W.~T.}\ \bibnamefont
  {de~Noronha}}, \bibinfo {author} {\bibfnamefont {A.~A.}\ \bibnamefont
  {Moreira}}, \bibinfo {author} {\bibfnamefont {N.~A.~M.}\ \bibnamefont
  {Ara{\'{u}}jo}}, \ and\ \bibinfo {author} {\bibfnamefont {J.~S.}\
  \bibnamefont {Andrade}},\ }\href {\doibase 10.1103/physrevresearch.2.043132}
  {\bibfield  {journal} {\bibinfo  {journal} {Physical Review Research}\
  }\textbf {\bibinfo {volume} {2}} (\bibinfo {year} {2020}),\
  10.1103/physrevresearch.2.043132}\BibitemShut {NoStop}%
\bibitem [{\citenamefont {Fehr}\ \emph {et~al.}(2009)\citenamefont {Fehr},
  \citenamefont {Jr}, \citenamefont {da~Cunha}, \citenamefont {da~Silva},
  \citenamefont {Herrmann}, \citenamefont {Kadau}, \citenamefont {Moukarzel},\
  and\ \citenamefont {Oliveira}}]{Fehr2009}%
  \BibitemOpen
  \bibfield  {author} {\bibinfo {author} {\bibfnamefont {E.}~\bibnamefont
  {Fehr}}, \bibinfo {author} {\bibfnamefont {J.~S.~A.}\ \bibnamefont {Jr}},
  \bibinfo {author} {\bibfnamefont {S.~D.}\ \bibnamefont {da~Cunha}}, \bibinfo
  {author} {\bibfnamefont {L.~R.}\ \bibnamefont {da~Silva}}, \bibinfo {author}
  {\bibfnamefont {H.~J.}\ \bibnamefont {Herrmann}}, \bibinfo {author}
  {\bibfnamefont {D.}~\bibnamefont {Kadau}}, \bibinfo {author} {\bibfnamefont
  {C.~F.}\ \bibnamefont {Moukarzel}}, \ and\ \bibinfo {author} {\bibfnamefont
  {E.~A.}\ \bibnamefont {Oliveira}},\ }\href {\doibase
  10.1088/1742-5468/2009/09/p09007} {\bibfield  {journal} {\bibinfo  {journal}
  {Journal of Statistical Mechanics: Theory and Experiment}\ }\textbf {\bibinfo
  {volume} {2009}},\ \bibinfo {pages} {P09007} (\bibinfo {year}
  {2009})}\BibitemShut {NoStop}%
\bibitem [{\citenamefont {Fehr}\ \emph {et~al.}(2012)\citenamefont {Fehr},
  \citenamefont {Schrenk}, \citenamefont {Araujo}, \citenamefont {Kadau},
  \citenamefont {Grassberger}, \citenamefont {Andrade},\ and\ \citenamefont
  {Herrmann}}]{Fehr2012}%
  \BibitemOpen
  \bibfield  {author} {\bibinfo {author} {\bibfnamefont {E.}~\bibnamefont
  {Fehr}}, \bibinfo {author} {\bibfnamefont {K.~J.}\ \bibnamefont {Schrenk}},
  \bibinfo {author} {\bibfnamefont {N.~A.~M.}\ \bibnamefont {Araujo}}, \bibinfo
  {author} {\bibfnamefont {D.}~\bibnamefont {Kadau}}, \bibinfo {author}
  {\bibfnamefont {P.}~\bibnamefont {Grassberger}}, \bibinfo {author}
  {\bibfnamefont {J.~S.}\ \bibnamefont {Andrade}}, \ and\ \bibinfo {author}
  {\bibfnamefont {H.~J.}\ \bibnamefont {Herrmann}},\ }\href
  {http://link.aps.org/doi/10.1103/PhysRevE.86.011117} {\bibfield  {journal}
  {\bibinfo  {journal} {Phys. Rev. E}\ }\textbf {\bibinfo {volume} {86}},\
  \bibinfo {pages} {011117} (\bibinfo {year} {2012})}\BibitemShut {NoStop}%
\bibitem [{\citenamefont {Schrenk}\ \emph {et~al.}(2012)\citenamefont
  {Schrenk}, \citenamefont {Araujo}, \citenamefont {Jr},\ and\ \citenamefont
  {Herrmann}}]{Schrenk2012}%
  \BibitemOpen
  \bibfield  {author} {\bibinfo {author} {\bibfnamefont {K.~J.}\ \bibnamefont
  {Schrenk}}, \bibinfo {author} {\bibfnamefont {N.~A.~M.}\ \bibnamefont
  {Araujo}}, \bibinfo {author} {\bibfnamefont {J.~S.~A.}\ \bibnamefont {Jr}}, \
  and\ \bibinfo {author} {\bibfnamefont {H.~J.}\ \bibnamefont {Herrmann}},\
  }\href {\doibase 10.1038/srep00348} {\bibfield  {journal} {\bibinfo
  {journal} {Scientific Reports}\ }\textbf {\bibinfo {volume} {2}} (\bibinfo
  {year} {2012}),\ 10.1038/srep00348}\BibitemShut {NoStop}%
\bibitem [{\citenamefont {Grassberger}(1999)}]{Grassberger1999}%
  \BibitemOpen
  \bibfield  {author} {\bibinfo {author} {\bibfnamefont {P.}~\bibnamefont
  {Grassberger}},\ }\href {\doibase 10.1016/s0378-4371(98)00435-x} {\bibfield
  {journal} {\bibinfo  {journal} {Physica A: Statistical Mechanics and its
  Applications}\ }\textbf {\bibinfo {volume} {262}},\ \bibinfo {pages} {251}
  (\bibinfo {year} {1999})}\BibitemShut {NoStop}%
\bibitem [{\citenamefont {Barth{\'{e}}l{\'{e}}my}\ \emph
  {et~al.}(1999)\citenamefont {Barth{\'{e}}l{\'{e}}my}, \citenamefont
  {Buldyrev}, \citenamefont {Havlin},\ and\ \citenamefont
  {Stanley}}]{Barthelemy1999}%
  \BibitemOpen
  \bibfield  {author} {\bibinfo {author} {\bibfnamefont {M.}~\bibnamefont
  {Barth{\'{e}}l{\'{e}}my}}, \bibinfo {author} {\bibfnamefont {S.~V.}\
  \bibnamefont {Buldyrev}}, \bibinfo {author} {\bibfnamefont {S.}~\bibnamefont
  {Havlin}}, \ and\ \bibinfo {author} {\bibfnamefont {H.~E.}\ \bibnamefont
  {Stanley}},\ }\href {\doibase 10.1103/physreve.60.r1123} {\bibfield
  {journal} {\bibinfo  {journal} {Physical Review E}\ }\textbf {\bibinfo
  {volume} {60}},\ \bibinfo {pages} {R1123} (\bibinfo {year}
  {1999})}\BibitemShut {NoStop}%
\bibitem [{\citenamefont {Barth{\'{e}}l{\'{e}}my}\ \emph
  {et~al.}(2000)\citenamefont {Barth{\'{e}}l{\'{e}}my}, \citenamefont
  {Buldyrev}, \citenamefont {Havlin},\ and\ \citenamefont
  {Stanley}}]{Barthelemy2000}%
  \BibitemOpen
  \bibfield  {author} {\bibinfo {author} {\bibfnamefont {M.}~\bibnamefont
  {Barth{\'{e}}l{\'{e}}my}}, \bibinfo {author} {\bibfnamefont {S.~V.}\
  \bibnamefont {Buldyrev}}, \bibinfo {author} {\bibfnamefont {S.}~\bibnamefont
  {Havlin}}, \ and\ \bibinfo {author} {\bibfnamefont {H.~E.}\ \bibnamefont
  {Stanley}},\ }\href {\doibase 10.1103/physreve.61.r3283} {\bibfield
  {journal} {\bibinfo  {journal} {Physical Review E}\ }\textbf {\bibinfo
  {volume} {61}},\ \bibinfo {pages} {R3283} (\bibinfo {year}
  {2000})}\BibitemShut {NoStop}%
\bibitem [{\citenamefont {BARTHELEMY}\ \emph {et~al.}(2003)\citenamefont
  {BARTHELEMY}, \citenamefont {BULDYREV}, \citenamefont {HAVLIN},\ and\
  \citenamefont {STANLEY}}]{BARTHELEMY2003}%
  \BibitemOpen
  \bibfield  {author} {\bibinfo {author} {\bibfnamefont {M.}~\bibnamefont
  {BARTHELEMY}}, \bibinfo {author} {\bibfnamefont {S.~V.}\ \bibnamefont
  {BULDYREV}}, \bibinfo {author} {\bibfnamefont {S.}~\bibnamefont {HAVLIN}}, \
  and\ \bibinfo {author} {\bibfnamefont {H.~E.}\ \bibnamefont {STANLEY}},\
  }\href {\doibase 10.1142/s0218348x03001689} {\bibfield  {journal} {\bibinfo
  {journal} {Fractals}\ }\textbf {\bibinfo {volume} {11}},\ \bibinfo {pages}
  {19} (\bibinfo {year} {2003})}\BibitemShut {NoStop}%
\bibitem [{\citenamefont {Paul}\ \emph {et~al.}(2000)\citenamefont {Paul},
  \citenamefont {Buldyrev}, \citenamefont {Dokholyan}, \citenamefont {Havlin},
  \citenamefont {King}, \citenamefont {Lee},\ and\ \citenamefont
  {Stanley}}]{Paul2000}%
  \BibitemOpen
  \bibfield  {author} {\bibinfo {author} {\bibfnamefont {G.}~\bibnamefont
  {Paul}}, \bibinfo {author} {\bibfnamefont {S.~V.}\ \bibnamefont {Buldyrev}},
  \bibinfo {author} {\bibfnamefont {N.~V.}\ \bibnamefont {Dokholyan}}, \bibinfo
  {author} {\bibfnamefont {S.}~\bibnamefont {Havlin}}, \bibinfo {author}
  {\bibfnamefont {P.~R.}\ \bibnamefont {King}}, \bibinfo {author}
  {\bibfnamefont {Y.}~\bibnamefont {Lee}}, \ and\ \bibinfo {author}
  {\bibfnamefont {H.~E.}\ \bibnamefont {Stanley}},\ }\href {\doibase
  10.1103/physreve.61.3435} {\bibfield  {journal} {\bibinfo  {journal}
  {Physical Review E}\ }\textbf {\bibinfo {volume} {61}},\ \bibinfo {pages}
  {3435} (\bibinfo {year} {2000})}\BibitemShut {NoStop}%
\bibitem [{\citenamefont {Melchert}\ and\ \citenamefont
  {Hartmann}(2007)}]{Melchert2007}%
  \BibitemOpen
  \bibfield  {author} {\bibinfo {author} {\bibfnamefont {O.}~\bibnamefont
  {Melchert}}\ and\ \bibinfo {author} {\bibfnamefont {A.~K.}\ \bibnamefont
  {Hartmann}},\ }\href {\doibase 10.1103/physrevb.76.174411} {\bibfield
  {journal} {\bibinfo  {journal} {Physical Review B}\ }\textbf {\bibinfo
  {volume} {76}} (\bibinfo {year} {2007}),\
  10.1103/physrevb.76.174411}\BibitemShut {NoStop}%
\bibitem [{\citenamefont {Dijkstra}(1959)}]{dijkstra1959}%
  \BibitemOpen
  \bibfield  {author} {\bibinfo {author} {\bibfnamefont {E.~W.}\ \bibnamefont
  {Dijkstra}},\ }\href@noop {} {\bibfield  {journal} {\bibinfo  {journal}
  {Numerische Mathematik}\ }\textbf {\bibinfo {volume} {1}},\ \bibinfo {pages}
  {269} (\bibinfo {year} {1959})}\BibitemShut {NoStop}%
\bibitem [{\citenamefont {Stauffer}(1994)}]{staufer1994}%
  \BibitemOpen
  \bibfield  {author} {\bibinfo {author} {\bibfnamefont {A.}~\bibnamefont
  {Stauffer}, \bibfnamefont {D.;~Aharony}},\ }\href@noop {} {\emph {\bibinfo
  {title} {Introduction to Percolation Theory}}}\ (\bibinfo  {publisher} {CRC
  Press},\ \bibinfo {year} {1994})\BibitemShut {NoStop}%
\bibitem [{\citenamefont {Schwarzer}\ \emph {et~al.}(1999)\citenamefont
  {Schwarzer}, \citenamefont {Havlin},\ and\ \citenamefont
  {Bunde}}]{Schwarzer1999}%
  \BibitemOpen
  \bibfield  {author} {\bibinfo {author} {\bibfnamefont {S.}~\bibnamefont
  {Schwarzer}}, \bibinfo {author} {\bibfnamefont {S.}~\bibnamefont {Havlin}}, \
  and\ \bibinfo {author} {\bibfnamefont {A.}~\bibnamefont {Bunde}},\ }\href
  {\doibase 10.1103/PhysRevE.59.3262} {\bibfield  {journal} {\bibinfo
  {journal} {Phys. Rev. E}\ }\textbf {\bibinfo {volume} {59}},\ \bibinfo
  {pages} {3262} (\bibinfo {year} {1999})}\BibitemShut {NoStop}%
\bibitem [{\citenamefont {Knackstedt}\ \emph {et~al.}(2002)\citenamefont
  {Knackstedt}, \citenamefont {Sahimi},\ and\ \citenamefont
  {Sheppard}}]{Knackstedt2002}%
  \BibitemOpen
  \bibfield  {author} {\bibinfo {author} {\bibfnamefont {M.~A.}\ \bibnamefont
  {Knackstedt}}, \bibinfo {author} {\bibfnamefont {M.}~\bibnamefont {Sahimi}},
  \ and\ \bibinfo {author} {\bibfnamefont {A.~P.}\ \bibnamefont {Sheppard}},\
  }\href {\doibase 10.1103/PhysRevE.65.035101} {\bibfield  {journal} {\bibinfo
  {journal} {Phys. Rev. E}\ }\textbf {\bibinfo {volume} {65}},\ \bibinfo
  {pages} {035101} (\bibinfo {year} {2002})}\BibitemShut {NoStop}%
\bibitem [{\citenamefont {Ara\'ujo}\ \emph {et~al.}(2005)\citenamefont
  {Ara\'ujo}, \citenamefont {Vasconcelos}, \citenamefont {Moreira},
  \citenamefont {Lucena},\ and\ \citenamefont {Andrade}}]{Araujo2005}%
  \BibitemOpen
  \bibfield  {author} {\bibinfo {author} {\bibfnamefont {A.~D.}\ \bibnamefont
  {Ara\'ujo}}, \bibinfo {author} {\bibfnamefont {T.~F.}\ \bibnamefont
  {Vasconcelos}}, \bibinfo {author} {\bibfnamefont {A.~A.}\ \bibnamefont
  {Moreira}}, \bibinfo {author} {\bibfnamefont {L.~S.}\ \bibnamefont {Lucena}},
  \ and\ \bibinfo {author} {\bibfnamefont {J.~S.}\ \bibnamefont {Andrade}},\
  }\href {\doibase 10.1103/PhysRevE.72.041404} {\bibfield  {journal} {\bibinfo
  {journal} {Phys. Rev. E}\ }\textbf {\bibinfo {volume} {72}},\ \bibinfo
  {pages} {041404} (\bibinfo {year} {2005})}\BibitemShut {NoStop}%
\bibitem [{\citenamefont {Barabasi}\ and\ \citenamefont
  {Stanley}(1995)}]{barabasi1995}%
  \BibitemOpen
  \bibfield  {author} {\bibinfo {author} {\bibfnamefont {A.}~\bibnamefont
  {Barabasi}}\ and\ \bibinfo {author} {\bibfnamefont {H.~E.}\ \bibnamefont
  {Stanley}},\ }\href@noop {} {\emph {\bibinfo {title} {Fractal Concepts in
  Surface Growth 1st Edition}}}\ (\bibinfo {year} {1995})\BibitemShut {NoStop}%
\bibitem [{\citenamefont {Tarjan}(1972)}]{Tarjan1972}%
  \BibitemOpen
  \bibfield  {author} {\bibinfo {author} {\bibfnamefont {R.}~\bibnamefont
  {Tarjan}},\ }\href {\doibase 10.1137/0201010} {\bibfield  {journal} {\bibinfo
   {journal} {SIAM Journal on Computing}\ }\textbf {\bibinfo {volume} {1}},\
  \bibinfo {pages} {146} (\bibinfo {year} {1972})}\BibitemShut {NoStop}%
\end{thebibliography}%

\end{document}